\documentclass[aps,preprint,superscriptaddress]{revtex4}

\usepackage{graphicx}
\usepackage{booktabs}
\usepackage{dcolumn}
\usepackage{bm}
\usepackage{color}
\usepackage{natbib}

\makeatletter

\renewcommand*{\p@subsection}{}
\renewcommand*{\p@subsubsection}{}
\makeatother
\newcounter{afn}

\begin{document}

\title{PULSED UCN PRODUCTION USING A DOPPLER SHIFTER AT J-PARC}

\author{S. Imajo}
\affiliation{Department of Physics, Kyoto University, Kyoto 606-8502, Japan}

\author{K. Mishima}
\affiliation{High Energy Accelerator Research Organization, Tsukuba, Ibaraki 305-0802, Japan}

\author{M. Kitaguchi}
\affiliation{Department of Physics, Nagoya University, Chikusa, Nagoya, 464-8602, Japan}

\author{Y. Iwashia}
\affiliation{Institute for Chemical Research, Kyoto University, Uji, Kyoto 611-0011, Japan}

\author{N. L. Yamada}
\affiliation{High Energy Accelerator Research Organization, Tsukuba, Ibaraki 305-0802, Japan}

\author{M. Hino}
\affiliation{Research Reactor Institute, Kyoto University, Kumatori, Osaka 590-0494, Japan}

\author{T. Oda}
\affiliation{Department of Nuclear Engineering, Kyoto University, Kyoto 615-8540, Japan}

\author{T. Ino}
\affiliation{High Energy Accelerator Research Organization, Tsukuba, Ibaraki 305-0802, Japan}

\author{H. M. Shimizu}
\affiliation{Department of Physics, Nagoya University, Chikusa, Nagoya, 464-8602, Japan}

\author{S. Yamashita}
\affiliation{International Center for Elementary Particle Physics, University of Tokyo, Bunkyo, Tokyo 113-0033, Japan}

\author{R. Katayama}
\affiliation{Department of Physics, The University of Tokyo, Bunkyo, Tokyo 113-0033, Japan}

\date{\today}

\begin{abstract}
We have constructed a Doppler-shifter-type pulsed ultra-cold neutron (UCN) source at the Materials and Life Science Experiment Facility (MLF) of the Japan Proton Accelerator Research Complex (J-PARC).
Very-cold neutrons (VCNs) with 136-$\mathrm{m/s}$ velocity in a neutron beam supplied by a pulsed neutron source are decelerated by reflection
on a {\it m=10} wide-band multilayer mirror, yielding pulsed UCN. The mirror is fixed to the tip of a 2,000-rpm rotating arm moving with 68-$\mathrm{m/s}$ velocity in the same direction as the VCN.
The repetition frequency of the pulsed UCN is $8.33~\mathrm{Hz}$ and the time width of the pulse at production is $4.4~\mathrm{ms}$.
In order to increase the UCN flux, a supermirror guide, wide-band monochromatic mirrors, focus guides, and a UCN extraction guide have been newly installed or improved.
The $1~\mathrm{MW}$-equivalent count rate of the output neutrons with longitudinal wavelengths longer than $58~\mathrm{nm}$ is $1.6 \times 10^{2}~\mathrm{cps}$, while that of the true UCNs is $80~\mathrm{cps}$.
The spatial density at production is $1.4~\mathrm{UCN/cm^{3}}$.
This new UCN source enables us to research and develop apparatuses necessary for the investigation of the neutron electric dipole moment (nEDM).

\end{abstract}


\maketitle

\section{\label{sec1}Introduction}

Neutrons with kinetic energies lower than the Fermi pseudo-potential of Ni are called ``ultra-cold neutrons (UCNs)."
The typical value of this potential is $243~\mathrm{neV}$, while the de Broglie wavelength, $\lambda$, of a UCN is more than $58~\mathrm{nm}$.
A UCN experiences a potential (which represents the neutron-nucleus interaction spatially averaged over millions of nuclei) having a form of a step function
which takes a nonzero value within the space where the UCN wave function overlaps the potential.
The incident wave-packets colliding with the averaged potential barrier that is higher than the kinetic energy of the neutron penetrate into the classically forbidden region
and hence some UCNs are scattered by nuclei inside the barrier, and then the wave-packets are almost fully reflected.
The magnetic dipole moment of a neutron in a magnetic field has a potential energy that is expressed by the scalar product of the neutron magnetic moment
$\mbox{\boldmath $\mu$}$ and the magnetic flux density $\mbox{\boldmath $B$}$, that is, $-\mbox{\boldmath $\mu$} \cdot \mbox{\boldmath $B$}$.
The absolute value corresponds to $60~\mathrm{neV}$ at a magnetic flux density of $1.0~\mathrm{T}$.
It means that polarized UCNs with velocities less than $3.4~\mathrm{m/s}$ cannot overcome the magnetic field.
As results of these features, UCNs can be stored in a magnetic bottle or a vessel with an interior surface which has a high material potential, and therefore
are useful for long-term storage experiments such as searches for the neutron electric dipole moment (nEDM)~\cite{lec01, lec02, lec03} or measurement of neutron lifetime~\cite{lec04, lec05}.
The non-zero nEDM implies the breaking of time-reversal invariance and its magnitude limits the predictions of some new physics models beyond the standard model of elementary particles.
The neutron lifetime determines the mass fraction of $^{4}$He in the early Universe and
the improvement of the accuracy of the lifetime verifies the consistency between the fractions resulted from lifetime measurements and satellite surveys~\cite{lec06, lec07}.
In addition, a UCN is useful probe in studies of a region far from the material surface, such as the behavior of the potential or the phonon dispersion relations~\cite{lec08}.

The most well-known UCN source is Steyerl neutron turbine at Institut Laue-Langevin (ILL)~\cite{lec09}.
Very cold neutrons (VCNs) generated in the reactor are guided by a slightly curved vertical neutron guide and decelerated below $50~\mathrm{m/s}$ by the gravitational field,
while thermal neutrons are filtered out by the curved guide.
The VCNs are guided into a vacuum chamber in which a turbine with 690 cylindrically curved blades rotates with a blade velocity of $25~\mathrm{m/s}$.
The blade surface is coated with Ni and the neutrons are multiply reflected by the moving blades.
Finally, the VCNs are decelerated below $6.2~\mathrm{m/s}$ by the Doppler effect.

A neutron Doppler shifter is a similar apparatus to the abovementioned turbine~\cite{lec10,lec11,lec12}.
In this apparatus, a small number of neutron mirrors move backwards at half the velocity of the target neutrons, and the neutrons are decelerated to the UCN region by a single reflection only.
The Doppler shifter is an effective device for decelerating pulsed neutrons with large velocity components in the guide direction.
The simple mechanism allows us to estimate the UCN flux precisely by a particle tracking simulation.
In the simplest case of the reflection from a rotating small flat reflector, there is a strong correlation between the velocities and recoil angles of decelerated neutrons.
The neutrons have a short-pulse time structure and therefore are useful for a time-of-flight (TOF) spectrometer~\cite{lec13}.
By using a shutter beside the source point whose open and close operation is synchronized with UCN production, the Doppler shifter can also be used for UCN storage experiments as if by using a steady source at the peak density of UCN~\cite{lec10,lec11}.
To the best of our knowledge, the first pulsed UCN source produced by a Doppler shifter was reported by Dombeck et al.~\cite{lec10,lec11,lec13};
this Doppler shifter was equipped with a package of Thermica crystals on the tip of a rotating arm and decelerated 400-$\mathrm{m/s}$ neutrons.

Another Doppler shifter has since been developed for research and development (R$\&$D) at the BL05 NOP beamline of the Materials and Life Science Experiment Facility (MLF) at the Japan Proton Accelerator Research Complex (J-PARC)~\cite{lec14, lec15}.
Neutron supermirrors~\cite{lec16, lec17, lec18} were selected for the reflector of our Doppler shifter because the reflectivity wavelength band is controllable freely.
The peak phase space density of 136-$\mathrm{m/s}$ VCN beam at J-PARC (in Ref.~\cite{lec19}) was 31 times greater than that of the UCNs at ILL PF2 (in Ref.~\cite{lec09}), hence the production of extremely high density UCN was expected.
The demonstration experiment of UCN production was carried out at a beam power of $120~\mathrm{kW}$.
As a result, a UCN count of $1.8~\mathrm{cps}$ was expected for a forthcoming 1-MW beam operation~\cite{lec19}.
The low count rate was due to a strong collimation of the injected VCNs and a small solid angle of UCN extraction.

We have proposed a nEDM search with a spallation UCN source (J-PARC P33)~\cite{lec20}.
Toward the realization of the plan, the Doppler shifter was used for the R$\&$D of an important apparatus for our nEDM search, which we named ``UCN rebuncher"~\cite{lec21}.
The UCN rebuncher is a neutron accelerator, which accelerates or decelerates pulsed UCNs suitably with the combination between the gradient magnetic field~\cite{lec22} and the adiabatic fast passage spin flipper~\cite{lec23}, rotates the UCN distribution in velocity-position phase space and then time-focuses the UCNs on any position.
The UCN rebuncher is necessary to improve the storage density of UCN at the nEDM experiment with the spallation source of low repetition frequency like J-PARC.
Because of such a mechanism of time-focus, the pulsed UCN produced by the Doppler shifter is suitable for the performance test of the UCN rebuncher.
In the test, UCNs extracted from the Doppler shifter were transported to a detector through Ni-coated UCN guides passing through the UCN rebuncher.
The focal length of the rebuncher was set to $3.8~\mathrm{m}$.
In consideration of an arrangement of the rebuncher at BL05, glass guides with total length of $5.6~\mathrm{m}$ were produced.
As a result of a UCN transport experiment, the 1 MW-equivalent total count rate at the end of the UCN guide was $0.1~\mathrm{cps}$ due to the low transmittance of the guide, while that of the environmental background was $0.03~\mathrm{cps}$.
In a time-focus experiment, unnecessary UCN pulses have to be interrupted by a synchronized shutter near the Doppler shifter in order to prevent frame overlap of the TOF and detect 3--4-m/s UCNs.
In that case the average flux of UCN is greatly thinned.
Therefore it was difficult to obtain the TOF spectrum with the shutter within a practical time span.
In order to perform the UCN rebuncher experiment at J-PARC/MLF BL05, it was necessary to increase the peak flux of the UCN pulse at least 10 times.

In order to increase the UCN flux, the incident VCNs should be transported into the Doppler shifter more efficiently. Therefore, we have newly installed supermirror guides, wide-band monochromatic mirrors, and focus guides in the neutron beamline.
Further, the produced UCNs should also be extracted without loss.
Thus, a Ni-coated UCN extraction guide has been inserted into the Doppler shifter.
The outcomes of these upgrades have been evaluated at beampowers of 200--$300~\mathrm{kW}$.
In this paper, we describe the details of the construction of our upgraded UCN source and its performance.
At present, the improved UCN rebuncher has been developing.
The performance test will be carried out by using the improved Doppler shifter.

\section{\label{sec2}PRINCIPLES OF UCN PRODUCTION USING DOPPLER SHIFTER}
In this section, we describe the principles of UCN production using a Doppler shifter.
As neutrons faster than UCNs can be reflected by Bragg scattering,
multilayer-structured materials such as mica crystals or artificial supermirrors can be applied to the Doppler shifter reflector. Hereafter, we call the reflector a ``Doppler mirror".

A neutron is reflected if it satisfies the Bragg condition in the frame of the moving Doppler mirror.
The velocity vector, $\mbox{\boldmath $v_{r}$}$, of the reflected neutron in the laboratory frame is given by
\begin{equation}
       \mbox{\boldmath $v_{r}$} = \mbox{\boldmath $v$} + 2 |(\mbox{\boldmath $v$}-\mbox{\boldmath $v_{m}$})\cdot \mbox{\boldmath $n$}|\mbox{\boldmath $n$},
       \label{equ01}
\end{equation}
where $\mbox{\boldmath $v$}$ is the velocity vector of the incident neutron, $\mbox{\boldmath $v_{m}$}$ is the velocity vector of the mirror surface at the reflection point of the neutron,
and $\mbox{\boldmath $n$}$ is the normal vector at the reflection point.
In the laboratory frame, the incident neutron receives a momentum stroke in the direction $\mbox{\boldmath $n$}$, as shown in equation (\ref{equ01})
The wavelength of the incident neutron is shifted to a shorter or longer value, as in the Doppler effect of a sound wave.
This is the reason why such devices are called Doppler shifters.

If $\mbox{\boldmath $v_{m}$}$ is close to $\mbox{\boldmath $v$}/2$, the reflected neutron is decelerated close to zero velocity.
The decelerated neutron is regarded as a UCN if the magnitude of $\mbox{\boldmath $v_{r}$}$ is less than $6.8~\mathrm{m/s}$.
Thus, a mechanism that can move a mirror in the same direction as and at the half-speed of the incident neutron is required, in order to form a Doppler shifter that decelerates an incident neutron to a UCN through a single reflection.
In the apparatus described here, continuous motion of the mirror is realized using a rotating arm, where the mirror is attached to the arm tip.

\begin{figure*}[htb]
  \centering
  \includegraphics*[width=150mm]{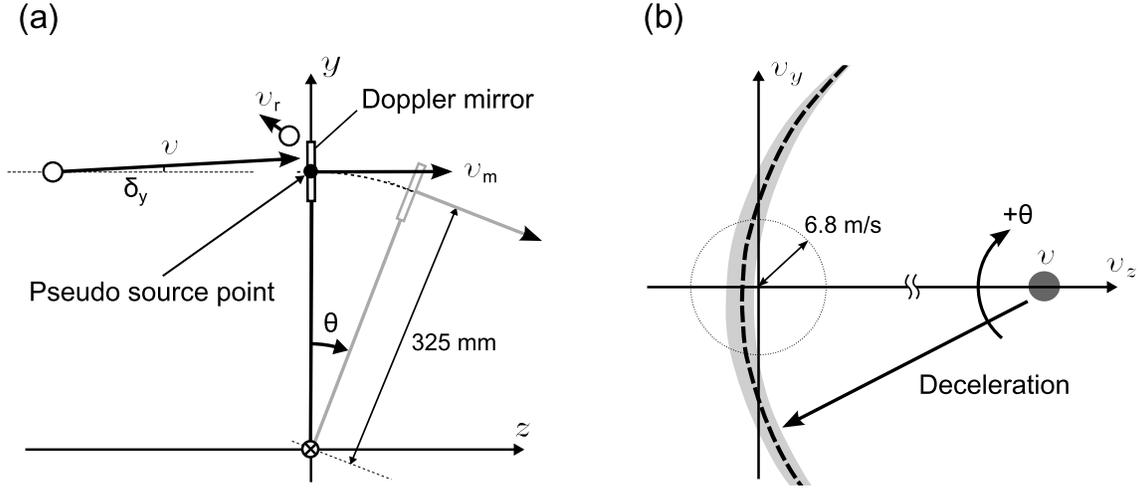}
  \caption{(a) Schematic diagram of Doppler shifter deceleration; (b) illustration of velocity phase space distribution of decelerated neutrons.\label{fig001}}
\end{figure*}

Next, we formulate the velocity of the decelerated neutron.
We define the beam direction as along the $z$-axis, the rotation axis of the arm is the $x$-axis,
and the axis perpendicular to the $x$- and $z$-axes is the $y$-axis, as shown in FIG. \ref{fig001}a.
The incident neutrons travel in the positive $z$ direction.
The Doppler mirror is positioned on the arm and the mirror surface is parallel to the $x$-axis and the direction of the arm.
We define the point where the beam crosses the Doppler mirror when the mirror surface comes to overlap the $y$-axis as the pseudo source point.
Then, the $x$, $y$, and $z$ components of the velocity of a reflected neutron, $v_{x}$, $v_{y}$, $v_{z}$, respectively, are given by
\begin{eqnarray}
       v_{x}&=&2 v_{m} \delta_{x},\label{equ02}\\
       v_{y}&=&2 v_{m}(\delta_{y}+\theta),\label{equ03}\\
       v_{z}&=&2 v_{m}-v+3 v_{m} \theta^{2},\label{equ04}
\end{eqnarray}
where $\delta_{x}$ and $\delta_{y}$ are the respective incident angles of the neutron in the $x$- and $y$-planes,
$v$ and $v_{m}$ are the magnitudes of $\mbox{\boldmath $v$}$ and $\mbox{\boldmath $v_{m}$}$, respectively,
and $\theta$ is the rotation angle of the mirror against the $y$-axis.
In these equations, we assume that $v_{m}$ is close to $v/2$
and $\delta_{x}$, $\delta_{y}$, and $\theta$ are small and ignore terms of order higher than the third order of their products.

In this case, the velocities of the decelerated neutrons are distributed on a parabola in the $v_{y}$-$v_{z}$ velocity space plane, as shown in FIG. \ref{fig001}b.
Then, $v_{z}$ is expressed as
\begin{equation}
       v_{z} = \frac{3}{4 v_{m}}(v_{y}-2 \delta_{y} v_{m})^{2} - (v - 2 v_{m}).
       \label{equ05}
\end{equation}

Suppose that $v = 136~\mathrm{m/s}$, $v_{m} = 68~\mathrm{m/s}$, $\delta_{x} = 0$, $\delta_{y} = 0$, the arm length is $325~\mathrm{mm}$ and, accordingly, the angular velocity of the rotation is $200\pi/3~\mathrm{rad/s}$.
Then, the condition for UCN production, $\sqrt{v_{y}^{2}+v_{z}^{2}} < 6.8~\mathrm{m/s}$, is expressed as $|\theta| < 50~\mathrm{mrad}$, from equations (\ref{equ03}) and (\ref{equ04}).
The angular width of $100~\mathrm{mrad}$ can be converted to a time width of $480~\mathrm{\mu s}$.

UCNs are produced in the vicinity of the pseudo source point.
As an increase of $\theta$ in $v_{y}$ can be canceled by non-zero $\delta_{y}$, as shown in equation (\ref{equ03}),
the actual width of $\theta$ for UCN production is wider for a divergent beam with a width of $\delta_{y}$. 
The estimated results yielded by a 3D Monte Carlo simulation based on ray tracing and incorporating the geometry of our Doppler shifter, which we refer to as the ``UCN simulation" in this paper, indicate that
$99.7\%$ of the extractable UCNs are produced within an angular width of $170~\mathrm{mrad}$
from an incident VCN beam with velocity divergence within $\pm 90~\mathrm{mrad}$.
Therefore, the time width of the extractable UCN production is $0.9~\mathrm{ms}$.

As the produced UCNs are extracted in the direction perpendicular to the orbit of the mirror, the time width of the produced UCN pulse is roughly identical to the mirror size.
This means that the maximum time width of the UCN pulse with 6.0-$\mathrm{m/s}$ velocity in the extraction direction is $5.0~\mathrm{ms}$, for the sample case of extraction in the upper direction with a 30-mm high mirror.
If UCNs are extracted along the $y$-axis in the UCN simulation,
the full width at half maximum (FWHM) of the UCN pulse time structure for a longitudinal velocity of approximately $6~\mathrm{m/s}$ is $2.9~\mathrm{ms}$ at production,
because of the incident beam divergence. The FWHM of the UCN pulse with a longitudinal velocity of more than $3.0~\mathrm{m/s}$ is $4.4~\mathrm{ms}$.
Therefore, the Doppler shifter with the design described above can generate sharply pulsed UCN of 4.4-ms width at production.

\section{\label{sec3}Experimental apparatus}
At the J-PARC MLF, pulsed proton beams are injected from the 3-GeV rapid cycling synchrotron ring (RCS) to the MLF source. Neutrons are then produced by spallation reactions and cooled by liquid hydrogen moderators.
The repetition frequency of the neutron production is $25~\mathrm{Hz}$.
The neutron production pauses for $360~\mathrm{ms}$ every 3 or 6 s, because proton beams are supplied to the 50-GeV main ring (MR)
when particle physics experiments in the Neutrino Experimental Facility or the Hadron Experimental Facility are being performed.

\begin{figure*}[htb]
  \centering
  \includegraphics*[width=160mm]{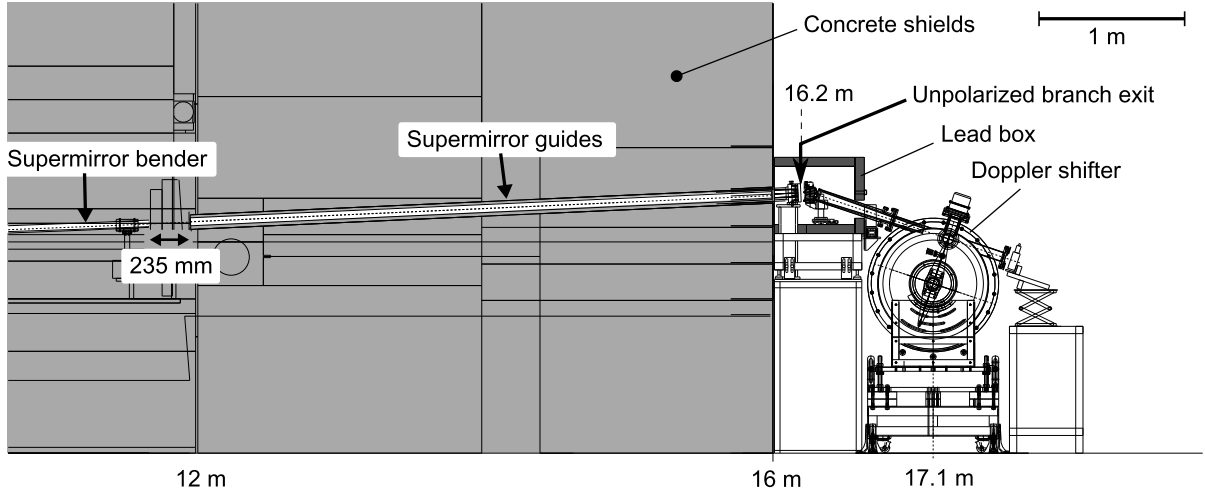}
  \caption{Layout of supermirror guides and Doppler shifter at J-PARC MLF BL05 with distance from the moderator.\label{fig002}}
\end{figure*}

At the BL05 port, an incident neutron beam is transported by $m=2$ supermirror guides, where $m$ is the critical angle normalized to that of Ni. The beam is then divided into three branches using supermirror bending guides: the polarized,
unpolarized, and low-divergence beam branches, which have been discussed previously in \cite{lec14,lec15}.
The Doppler shifter is positioned on the unpolarized beam branch as shown in FIG. \ref{fig002},
where the neutrons are guided upwards at $2.6^{\circ}$ by a $m = 3$ arc-shaped supermirror bender of 100-m radius and 4.0-m length~\cite{lec14}.
The bending guide has a cross section of $40~\mathrm{mm~(width)} \times 50~\mathrm{mm~(height)}$, which is separated into five channel structures of $40~\mathrm{mm~(width)} \times 10~\mathrm{mm~(height)}$.
Fast neutrons escape from the guide at the curve and only cold neutrons are transported.
Neutrons transported through the bender pass $m=3$ supermirror guides to reach the exit of the concrete shielding surrounding the apparatus.
These guides have 4.2-$\mathrm{m}$ length and the internal cross section is $40~\mathrm{mm~(width)} \times 50~\mathrm{mm~(height)}$.
The gap between the bender and guide is $235~\mathrm{mm}$ wide, and is designed to be as short as possible to transport VCN with large divergence.
Note that a B$_{4}$C collimator was used instead of the supermirror guide tube when the demonstration experiment discussed in this paper was performed~\cite{lec19}.

We measured the fluxes of the unpolarized branch with the supermirror guides or the B$_{4}$C collimators in terms of wavelength distributions using a neutron beam monitor, MNH10/4.2F~\cite{lec24}, with a collimator of $\phi 10~\mathrm{mm}$.
The detector was mounted on the unpolarized branch exit at the 16.2-m position shown in FIG. \ref{fig002}.
The neutron pulse interval at MLF is usually $40~\mathrm{ms}$, so neutrons with velocities less than $400~\mathrm{m/s}$ cannot normally be distinguished at the unpolarized branch.
Therefore, neutrons produced by the last beam pulse immediately before the beam supply to the MR were selected from the data to prevent frame overlap with slower neutrons,
and the TOFs from their production were then converted to the corresponding wavelengths.
The spectra are shown in FIG. \ref{fig003}, which were calculated using detection efficiency compensation and normalized to the designed beampower of $1~\mathrm{MW}$.

\subsection{\label{sec31}Supermirror guides}
\begin{figure*}[htbp]
  \centering
  \includegraphics*[width=120mm]{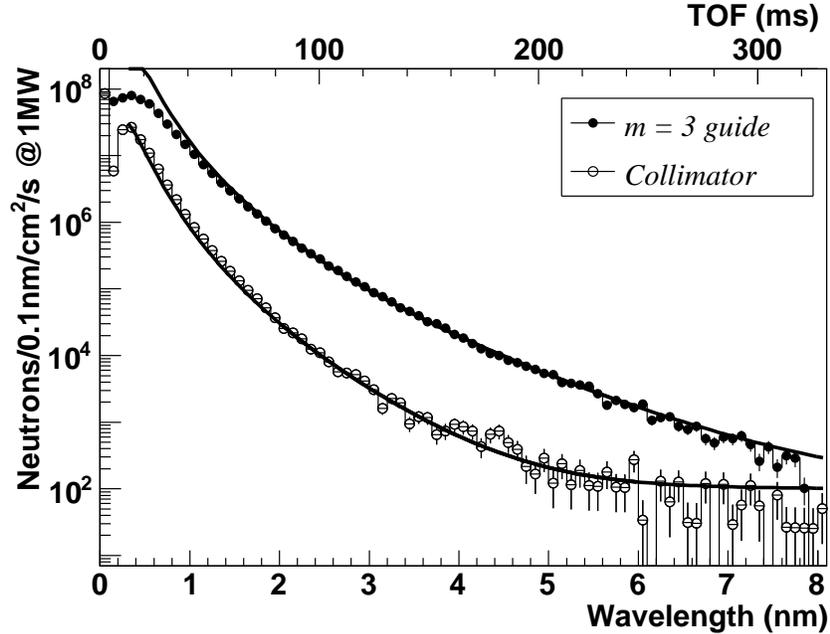}
  \caption{Neutron fluxes of unpolarized branch at J-PARC MLF BL05. The closed and open circles indicate the fluxes obtained with the $m=3$ supermirror guides and the B$_{4}$C collimator, respectively.\label{fig003}}
\end{figure*}

It is possible to fit the flux data sets using a Boltzmann distribution under certain conditions, where
\begin{equation}
       \phi(\lambda)= \Phi \frac{4 \pi h^{3}}{\left( 2 \pi m k_{B} \right)^{3/2}}\frac{1}{T^{3/2}~\lambda^4}\exp \left(-\frac{h^{2}}{2 m k_{B}} \frac{1}{T~\lambda^2} - a \lambda \right) + C_{b}.
       \label{equ06}
\end{equation}
In this equation, $\phi (\lambda)$ is the flux distribution, $\Phi$ is the total flux, $h$ is the Planck constant, $m$ is the neutron mass, $k_{B}$ is the Boltzmann constant,
$T$ is the neutron temperature, $a$ is the attenuation coefficient (such as absorption by the Al windows),
and $C_{b}$ is the neutron environmental background.
This expression can be applied to cases where the neutron divergence is independent of wavelength,
such that the neutrons are perfectly reflected or absorbed by a neutron guide or collimator. The flux is primarily restricted by the solid angle of the guide entrance or the collimator exit.
In the case of BL05, the $m=2$ supermirror guides with incident angle less than the critical angle have more than $90\%$ reflectivity; hence, this condition is roughly satisfied for low-divergence VCNs.

In a divergence measurement with a 1-mm-diameter pinhole and a position-sensitive detector~\cite{lec25,lec26},
the FWHM of the $x'$- and $y'$-divergence of VCNs with wavelengths longer than $2.0~\mathrm{nm}$ ($198~\mathrm{m/s}$)
were found to be 50 and $54~\mathrm{mrad}$ respectively, where the $x'$- and $y'$-axes were the horizontal and vertical directions on the plane parallel to the unpolarized branch window.
The divergence of the $99\%$ VCNs was within $\pm 60~\mathrm{mrad}$ in each direction.
Because the critical velocity of the $m=2$ supermirror at zero incident angle is $13.6~\mathrm{m/s}$,
a VCN with a glancing angle of $60~\mathrm{mrad}$ and $v < 13.6/ \tan(0.06) = 226~\mathrm{m/s}$ ($1.74~\mathrm{nm}$) is perfectly reflected by the surface.
Accordingly, the flux attenuation for such a VCN is almost independent of its wavelength in the transport, excluding the case of neutron capture in Al.
Therefore, the spectra over $2.0$-$\mathrm{nm}$ wavelength shown in FIG. \ref{fig003} are described by the Boltzmann distribution of equation (\ref{equ06}).
In the fitting, we assumed the existence of a constant $C_{b} = 1000~\mathrm{neutrons/nm}$,
because the slopes of the histograms in FIG. \ref{fig003} differed; this was most likely due to background neutrons.

Based on the fit observed between $2.0$- and $6.0$-$\mathrm{nm}$ wavelength for the data obtained with the supermirror guide, the most probable values of $a$ and $\Phi$ were $0.448 \pm 0.006 ~\mathrm{nm^{-1}}$ and $2.0 \times 10^{9}~\mathrm{neutrons / cm^{2}/s}$, respectively.
The value of $T$ was fixed to $57.3~ \mathrm{K}$, which was calculated via simulation of the moderator~\cite{lec27,lec28}.
Note that the flux around the 2.9-nm (136-m/s) components increased 30-fold compared with the fit results for the supermirror guide and B$_{4}$C collimator.

\subsection{\label{sec32}VCN transportation}
\begin{figure*}[htbp]
  \centering
  \includegraphics*[width=130mm]{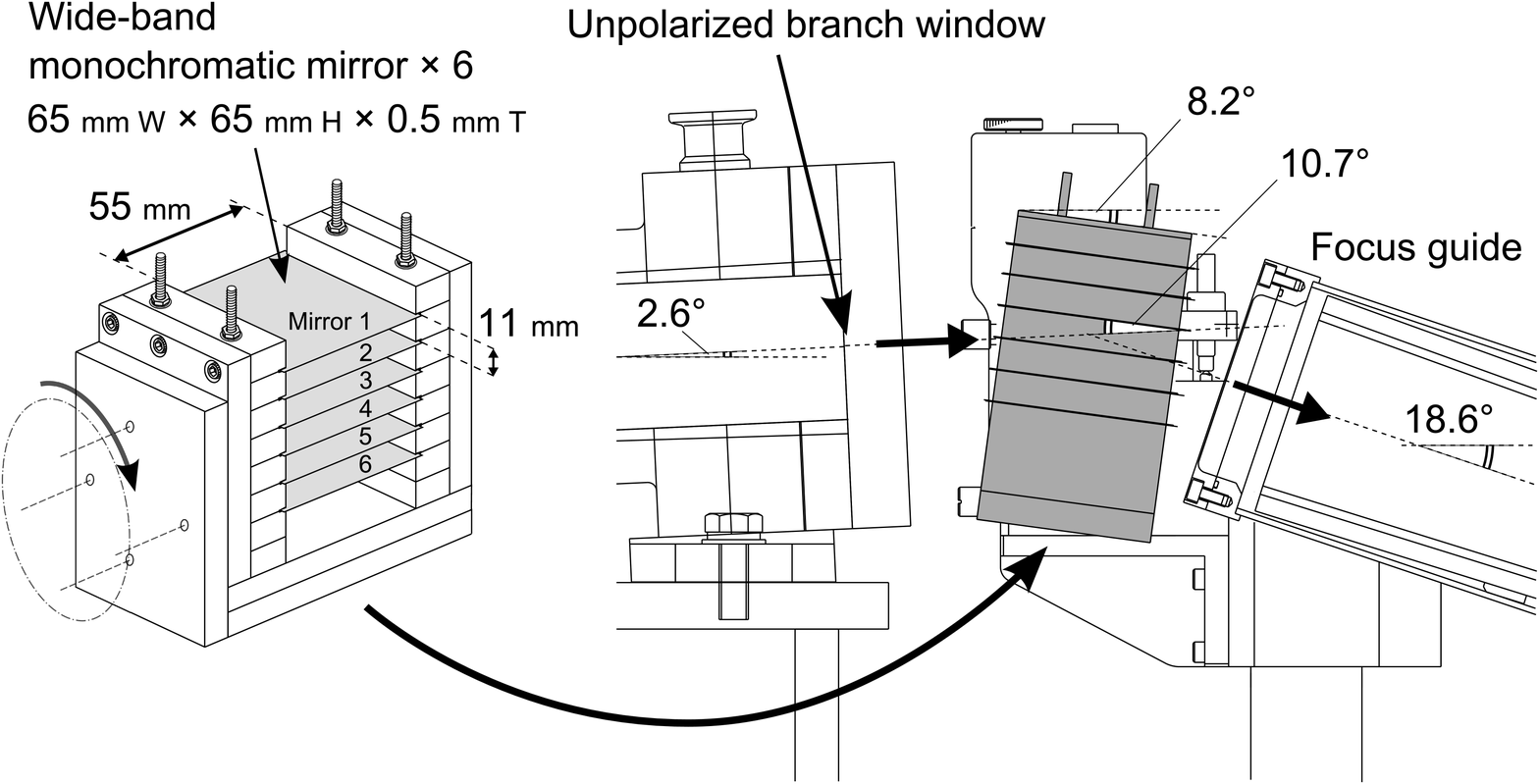}
  \caption{Arrangement of wide-band monochromatic mirrors. \label{fig004}}
\end{figure*}

\begin{figure*}[htbp]
  \centering
  \includegraphics*[width=120mm]{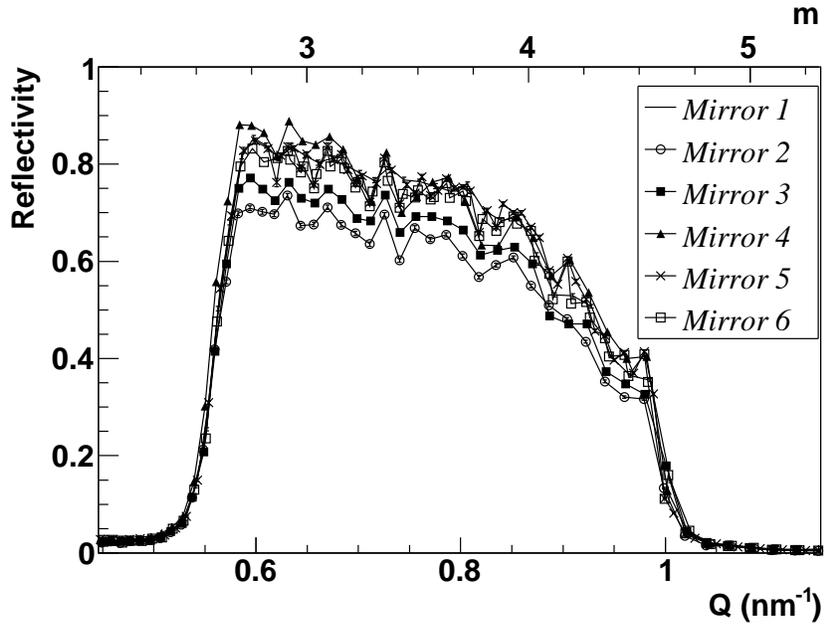}
  \caption{Reflectivity of wide-band monochromatic mirrors.
  \label{fig005}}
\end{figure*}

Six $m = 3.5$ wide-band multilayer monochromatic mirrors, fabricated at the Kyoto University Research Reactor Institute (KURRI), were installed
in front of the unpolarized branch, as shown in FIG. \ref{fig004}. This was to ensure that only useful neutrons for UCN production were extracted, and also to decrease the background radiation.
The reflectivities of the monochromatic mirrors, as measured at SOFIA (BL16) at J-PARC~\cite{lec29,lec30}, are shown in FIG. \ref{fig005}.
These values are plotted as a function of neutron momentum transfer $Q=(4 \pi/ \lambda) \sin \theta$, where $\theta$ is the glancing angle of an incident neutron.
As shown in FIG. \ref{fig004}, the direction of the VCN extraction is set to an angle of $18.6^\circ$ below the horizontal.
Hence, the mirrors are required to be arranged at an angle of $(18.6^\circ+2.6^\circ)/2-2.6^\circ=8.0^\circ$.
$Q$ is calculated as $0.80~\mathrm{nm^{-1}}$ for a VCN with $2.9$-$\mathrm{nm}$ wavelength making contact with the mirror at $\theta = 10.6^\circ$,
hence, the reflectivities are designed to be distributed around $Q=0.80$.
Accordingly, VCNs with approximately $2.9$-$\mathrm{nm}$ wavelength are reflected downward, injected into the focus guide and then guided into the Doppler shifter, as will be described below.

The mirror substrates in this apparatus are composed of 0.5-mm-thick Si with dimensions of $65~\mathrm{mm~(width)} \times 65~\mathrm{mm~(height)}$ and are arranged with 11-mm spacing in order to cover the beam cross section within the shortest path.
The mirror angle was fine-tuned in order to maximize the UCN yield, with the mirrors being tilted off the horizontal by $8.2^\circ$.
Therefore, the beam makes contact with the monochromatic mirrors at a glancing angle of $10.7^\circ$. 
Note that the mirrors are contained within a 5-cm-thick lead box covered by a 5-mm-thick B$_{4}$C region, as shown in FIG. \ref{fig002}, to reduce the background noise from neutrons and gamma rays.

\begin{figure*}[htbp]
  \centering
  \includegraphics*[width=150mm]{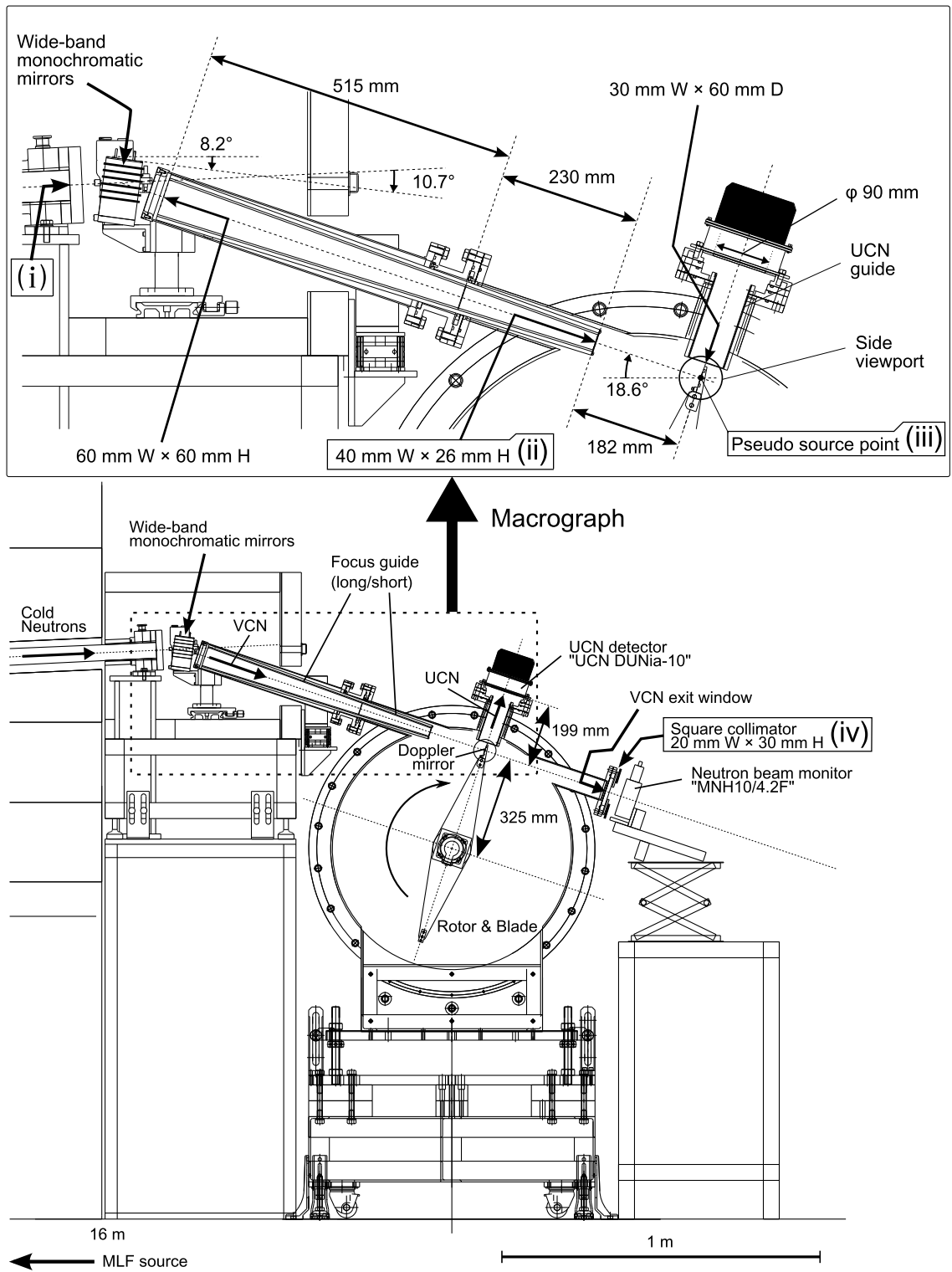}
  \caption{Doppler shifter and peripheral device layout. The inset is a macrograph of the wide-band monochromatic mirror, focus guide, and UCN extraction guide arrangement. \label{fig006}}
\end{figure*}

As shown in FIG. \ref{fig006}, the VCNs reflected by the monochromatic mirrors are guided into the Doppler shifter by the Ni-coated guides, which consist of Ni-evaporated trapezoidal glasses.
The guide cross section narrows towards the exit, so that the VCN flux is concentrated on the pseudo source point, while the divergence increases correspondingly.
The guides are composed of two parts: a long ($515~\mathrm{mm}$) and short ($230~\mathrm{mm}$) guide.
The cross section of the VCN entrance is $60~\mathrm{mm~(width)} \times 60~\mathrm{mm~(height)}$ and that of the exit is $40~\mathrm{mm~(width)} \times 26~\mathrm{mm~(height)}$.
The exit size was optimized to maximize the UCN yield using a Monte Carlo simulation.
The end of the short guide is positioned $182~\mathrm{mm}$ from the pseudo source point.
The TOF distance for the transit between the moderator surface and the pseudo source point is $17.2~\mathrm{m}$,
therefore, a VCN with 136-m/s velocity arrives at the pseudo source point $127~\mathrm{ms}$ after its generation.

\begin{figure*}[htbp]
  \centering
  \includegraphics*[width=120mm]{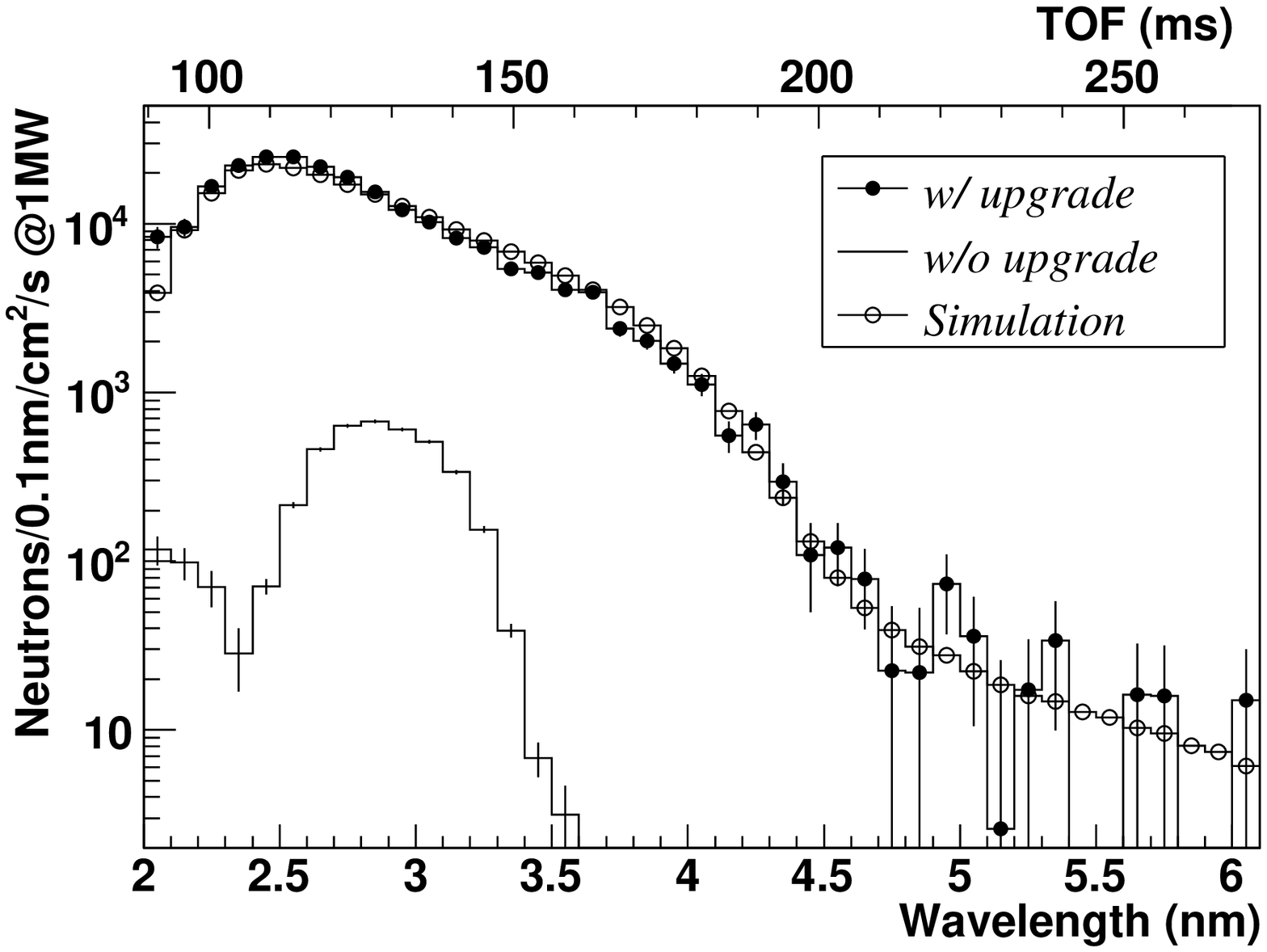}
  \caption{VCN flux reflected by monochromatic mirrors with and without upgrade, and simulated flux with upgrade,
  at Doppler shifter VCN exit window.
  Each data set is compensated based on each detector efficiency and scaled to a beampower of $1~\mathrm{MW}$. \label{fig007}}
\end{figure*}

FIG. \ref{fig007} shows the VCN flux at the exit window downstream of the Doppler shifter, as measured by the
neutron beam monitor MNH10/4.2F with a square collimator of $20~\mathrm{mm~(width)} \times 30~\mathrm{mm~(height)}$ (see FIG. \ref{fig006}).
The data sets were compensated based on the detector efficiency in order to convert the neutron flux.

The closed circles indicate the flux obtained under the present conditions, with the supermirror, Ni-coated guides, and 6 wide-band monochromatic mirrors equipped,
while the line without circles indicates the flux measured in the demonstration experiment~\cite{lec19}, where only the B$_{4}$C collimator and 4 narrow-band monochromatic mirrors were installed.
The two plots are scaled to a beampower of $1~\mathrm{MW}$.
We analyzed the last and succeeding time frames before the MR injections using the same approach described in section \ref{sec31}.
Then, the frame overlaps of the preceding pulse were subtracted, assuming that the preceding pulse had the same tail as the final pulse.

The open circles in FIG. \ref{fig007} indicate a VCN flux calculated with
the VCN transport after the beam exit of the unpolarized branch using the UCN simulation.
In this simulation, the initial $x'$- and $y'$-values ($x'$ and $y'$ are defined in section \ref{sec31}) were given uniformly, and the velocity divergences in the $x'$- and $y'$-directions were determined using the result of the pinhole measurement mentioned in section \ref{sec31}.
The velocities in the branch direction were given according to the probability distribution in equation (\ref{equ06}), $\phi(\lambda)/\Phi$.
The pulse shape and the delay by moderation of $0.10~\mathrm{ms}$ listed in~\cite{lec28} were taken into account in the UCN simulation. 
The monochromatic mirrors with the reflectivities shown in FIG. \ref{fig005} and the focus guide were implemented.
For comparison of the spectrum shape, the simulated count integrated between $2.5$ and $6.0~\mathrm{nm}$ was normalized to the measured count in the same region, where the overlaps by the preceding pulses were well removed.
The simulated spectrum agreed well with the measurement.

The $1~\mathrm{MW}$-equivalent simulated VCN fluxes $ \phi_{s}$ with wavelength range of 2.8--$3.0~\mathrm{nm}$, which is the target region for the Doppler shifter, at the position (i--iv)
in FIG. \ref{fig006} and measured fluxes $ \phi_{m}$ at the position (i) and (iv) are summarized in TABLE. \ref{tab01}.
The meaning of each position number is as follows: (i): the unpolarized branch window, (ii): the exit of the focus guide, (iii): the pseudo source point, and (iv): the square collimator on the VCN exit window.
The cross sections $ S_{c}$ used in the flux calculation at each position are also summarized for reference.
The value of $ \phi_{m}$ at the position (i) was estimated by a numerical integration of equation (\ref{equ06}).
$^{\dag}$The value of $ \phi_{s}$ at the position (i) was normalized by $ \phi_{m}$ at the same position.
The decrease of $ S_{c} \phi_{s}$ from the position (i) to (ii) was due to the the transport efficiency of the monochromatic mirrors and the focus guide,
which were estimated to be $64\%$ and $84\%$ for 2.9-nm VCN by the UCN simulation, respectively.
The decrease of flux at the downstream of the position (ii) was due to beam diffusion in the free space of the Doppler shifter.
In the estimation with the ratio of $ \phi_{s}$ at the position (iii) to (iv), the measured flux at the position (iii) was found to be $8.7 \times 10^{4}~\mathrm{neutrons/cm^{2}/s}$, which is 43-fold that of the demonstration experiment.
The presumed flux, however, was half the simulated value at the same position.
The loss is suspected to be due to the wide-band monochromatic mirrors and/or the focus guide.
We are currently investigating the cause of the large flux decrease observed in the experiment.

\begin{table}[htbp]
  \caption{Change of $1~\mathrm{MW}$-equivalent simulated VCN flux $ \phi_{s}$ and measured flux $ \phi_{m}$ with wavelength range of 2.8--$3.0~\mathrm{nm}$
  in cross section $S_{c}$ at the position (i--iv) in FIG. \ref{fig006}.
  $^{\dag}$The value of $ \phi_{s}$ at the position (i) was normalized by the measured one.}
  \label{tab01}
  \begin{center}

    \begin{tabular}{c c c c}
      \hline
      Position ~~& $S_{c}~\mathrm{[cm^{2}]}$~~ & $ \phi_{s}~\mathrm{[cm^{-2}s^{-1}]}$~~ & $\phi_{m}~\mathrm{[cm^{-2}s^{-1}]}$~~\\
      \hline
      (i)~~& $4.0 \times 5.0$~~ & $^{\dag}2.3 \times 10^{5}$~~ & $2.3 \times 10^{5}$~~\\
      (ii)~~& $4.0 \times 2.6$~~ & $2.3 \times 10^{5}$~~ & -~~\\
      (iii)~~& $3.0 \times 3.0$~~ & $1.8 \times 10^{5}$~~ & -~~\\
      (iv)~~& $2.0 \times 3.0$~~ & $5.6 \times 10^{4}$~~ & $2.8 \times 10^{4}$~~\\
      \hline
    \end{tabular} 
  \end{center}
\end{table}

\subsection{\label{sec33}Doppler shifter}
The structure and setup of the Doppler shifter are shown in FIG. \ref{fig006}.
A blade is rotated like the hand of a clock by a servomotor in a cylindrical vacuum chamber.
The vacuum around the motor shaft is maintained using magnetic fluid seals.
A mirror holder is located at the tip of the blade and four neutron mirrors, which form the Doppler mirror, are stacked and mounted on the holder.
The mirrors consist of $d=2.9$-$\mathrm{nm}$ wide-band multilayers with $2.6$--$3.2$-$\mathrm{nm}$ d-spacings~\cite{lec16,lec17,lec18},
and each mirror has a different reflectivity distribution.
Each mirror has ion-beam-sputtered NiC/Ti multilayers on both surfaces of 0.3-mm-thick Si wafer~\cite{lec16}, which was created at KURRI.
The mirror size is $35~\mathrm{mm~(width)} \times 30~\mathrm{mm~(height)}$.
Further, 2.5-$\mathrm{mm}$-wide areas on both sides are covered by stoppers; hence, the effective reflection area of the Doppler mirror is $30~\mathrm{mm} \times 30~\mathrm{mm}$.

\begin{figure*}[htbp]
  \centering
  \includegraphics*[width=120mm]{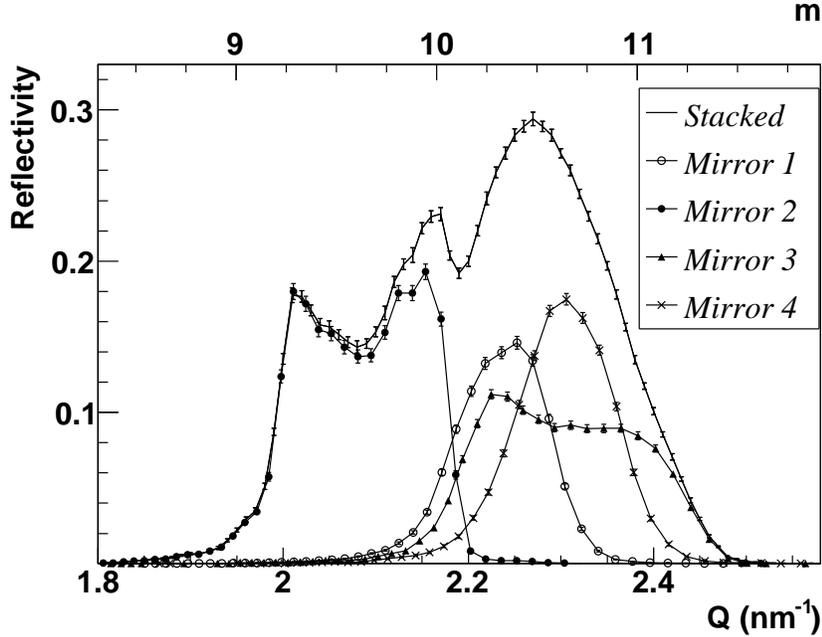}
  \caption{Reflectivities of four $m=10$ wide-band multilayer mirrors and numerically calculated reflectivities of the four stacked mirrors. \label{fig008}}
\end{figure*}

FIG. \ref{fig008} shows the reflectivities of the four Doppler mirrors as functions of wavelength measured at SOFIA,
along with those of the stacked mirrors, which were estimated using a numerical calculation considering multiple reflections.
The mirrors were set on the mirror holder from the bottom to surface in descending order of reflectivity.
The reflectivity at $Q = 2.16$, which corresponds to $68.0~\mathrm{m/s}$, was $0.229 \pm 0.004$.

The rotating radius of the mirror center is $325~\mathrm{mm}$ and, in the experiment described here, the rotation frequency was set to $2,000~\mathrm{rpm}$.
The center of the mirror moves with a velocity of $68~\mathrm{m/s}$ and the mirror interacts with incoming 25-Hz VCNs of 136-m/s velocity once in every three incident pulses.
Thus, the UCN production repetition frequency is $8.33~\mathrm{Hz}$.
As a UCN of 3-m/s velocity travels a distance of $300~\mathrm{mm}$ in $100~\mathrm{ms}$, this production frequency is suitable for the demonstration of UCN production with TOF measurements.

The 12-MHz master clock of J-PARC and the 25-Hz continuous timing pulse are fed to a motor control unit and divided pulses are used to synchronize the mirror rotation with the VCN incidence.
The phase offset of the rotation is controlled by the control unit in real time.
The phase control jitter is maintained at less than $\pm 3.5~\mathrm{mrad}$.

The Doppler shifter has three UCN extraction ports, with one above and two beside the pseudo source point.
As the UCN simulation showed that the decelerated neutrons are distributed on the parabola of equation (\ref{equ05}),
as shown in FIG. \ref{fig001}b, the UCN detector was mounted on the upper port in this study, as shown in FIG. \ref{fig006}.
The two side ports functioned as viewports for measurement of the rotating mirror timing using a pair of a laser and a photo-sensor,
which indicated the timing when the mirror holder passed the pseudo source point.
The accuracy of the timing detection was within $\pm 0.1 ~\mathrm{ms}$.

A UCN guide was inserted into the upper extraction port, which is connected from $26~\mathrm{mm}$ above the center of the Doppler mirror to 30 mm in front of the UCN detector, in order to collect UCNs with a wide solid angle.
The interior dimensions of the guide were $30~\mathrm{mm~(width)} \times 60~\mathrm{mm~(depth)}$ and the height was $143~\mathrm{mm}$.
Ni-coated silicon wafers were glued on the guide inner surface.
The TOF distance between the center of the Doppler mirror and the UCN detector window was $199~\mathrm{mm}$.
The rotation plane was aligned along the center line of the unpolarized branch.
The vacuum chamber can rotate around the motor axis from $-5^\circ$ to $30^\circ$ to match the angle of incoming VCNs.
The angle was set to $18.6^\circ$ upward.
During operation, the vacuum chamber was maintained at less than 3 Pa using a dry pump.

\subsection{\label{sec34}UCN detector}
For UCN detection, we used a proportional gas counter, UCN DUNia-10, manufactured by A. V. Strelkov.
The cylindrical gas chamber was filled with $^{3}$He gas at $2.7\times 10^{3}~\mathrm{Pa}$, and then with a gas mixture of $1\%$ CH$_4$ and $99\%$ Ar up to $1.1\times 10^{5}~\mathrm{Pa}$.
The detector is sensitive to slow neutrons only, as a result of the rarefied $^{3}$He gas.
The chamber is $50~\mathrm{mm}$ in length and has an entrance window of $90~\mathrm{mm}$ in diameter placed at the bottom.
The entrance window is covered with a pure Al foil of 0.1-mm thickness, and the foil is supported by metal beams on the outside of the chamber, for use in vacuum.
In the chamber, a W anode wire is stretched in the radial direction, with a high applied voltage of $1.0~\mathrm{kV}$.
Note that we replaced the built-in DUNia-10 preamplifier with one that is insensitive to the microphonic noise caused by the mechanical vibrations of the motor rotation.
Further, neutron output signals of $+2~\mathrm{V}$ in height and $5~\mathrm{\mu s}$ in width are now input directly into a 16 ch input PHA+LIST system, 760-PRU06PIK, and recorded, along with the proton-delivery timing signals and the $1/3$-divided 25-Hz pulses for the TOF analysis.

The detection efficiency was evaluated by comparing the wavelength spectra measured for the cold neutrons
by the UCN detector and those measured by a 1-MPa $^{3}$He proportional counter, RS-P4-0812-223, of 25.4-mm diameter with a 0.5-mm-thick stainless steel wall.
Here, we label the attenuation by the Al window and the stainless steel wall as $\tau_{A}$ and $\tau_{B}$, respectively, and the detection efficiency of the UCN detector
and the 1-MPa counter are $\epsilon_{A}$ and $\epsilon_{B}$, respectively.
Then, the detection ratio of the two detectors, $r( \lambda )$, is given by
\begin{equation}
       r( \lambda ) = \frac{\tau_{A} \epsilon_{A}}{\tau_{B} \epsilon_{B}}=\frac{\exp (-p_{0} \lambda) (1-\exp (-p_{1} \lambda))}{\exp (-p_{2} \lambda) (1-\exp (-p_{3} \lambda))},
       \label{equ07}
\end{equation}
where $p_{0}$ is the attenuation coefficient of the Al window, $p_{1}$ is the reaction coefficient of the $^{3}$He gas of the UCN detector,
$p_{2}$ is the attenuation coefficient of the stainless steel wall and $p_{3}$ is the reaction coefficient of the $^{3}$He gas of the 1-MPa counter.
The values of $p_{0}$ and $p_{2}$ were calculated to be $9.1 \times 10^{-4}$ and $6.0 \times 10^{-2}~\mathrm{nm^{-1}}$, respectively,
using the total Al cross section at $298~ \mathrm{K}$ in Ref. \cite{lec31} and the absorption cross section of Fe, instead of stainless steel, in Ref. \cite{lec32}.
Through fitting in the $0.50$--$1.04$-$\mathrm{nm}$ region, 
$p_{1}$ was evaluated as $(1.001 \pm 0.004) \times 10^{-1}~\mathrm{nm^{-1}}$.
The result was within $5\%$ agreement with $9.5 \times 10^{-2}~\mathrm{nm^{-1}}$, which was calculated from the $^{3}$He gas conditions of $2.7\times 10^{3}~\mathrm{Pa}$ and 50-mm thickness at $300~ \mathrm{K}$.
Further, $p_{3}$ was estimated as $7.7 \pm 0.6~\mathrm{nm^{-1}}$.
The detection efficiency decreased by $20\%$ when the incident point was 35-mm from the wire.
This discrepancy was caused by insufficient collection of the electrons by the anode wire,
as the pulse height was reduced by $25\%$ in the region far from the wire.

As a result, the detection efficiency of the UCN detector without attenuation in the Al foil, $\epsilon (\lambda)$, can be expressed as
\begin{equation}
       \epsilon (\lambda) = 1-\exp (-0.100 \lambda),
       \label{equ08}
\end{equation}
with $\lambda$ in nanometers.

\section{\label{sec4}MEASUREMENT OF DOPPLER-SHIFTED NEUTRONS}
In this section, we describe the results of UCN production experiments and the analysis using the UCN simulation.

\subsection{\label{sec41}Doppler-shifted neutrons}
\begin{figure*}[htbp]
  \centering
  \includegraphics*[width=150mm]{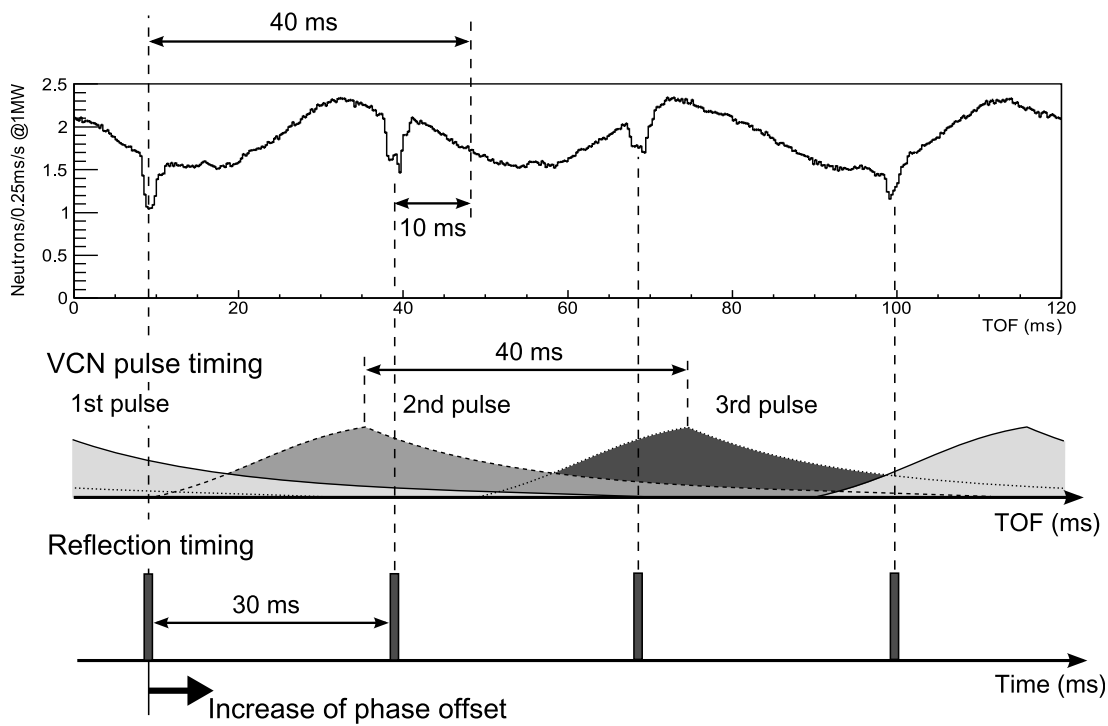}
  \caption{TOF spectra of VCNs passing through Doppler shifter (top). The four dips show the overlap of the Doppler mirror with the VCN beams.
  The middle and lower parts of this figure show the timing relation between the VCN pulse incidence and the VCN reflection by the Doppler mirror, schematically.\label{fig009}}
\end{figure*}

Typical TOF spectra for the VCN measured downstream of the operating Doppler shifter by the MNH10/4.2F neutron monitor are shown in the upper part of FIG. \ref{fig009}.
Because the full time width of a VCN pulse separated by the wide-band monochromatic mirrors was approximately $100~\mathrm{ms}$, the spectra appear to be continuous because of the frame overlap.
As shown in the middle and lower parts of FIG. \ref{fig009}, the mirror rotates four times during the three VCN pulse incidences, so that four dips of 30-ms intervals appear every three $40$-$\mathrm{ms}$ pulses.
These dips were used for a cross-check of the VCN reflection timing and an estimation of the reflected VCN velocity.
The calculated error of the timing estimation was $\pm 0.2~\mathrm{ms}$.
In the experiment, the phase offsets estimated using this method were consistent with those evaluated by the photo-sensor, as mentioned in section \ref{sec33}.
The timings obtained by these two methods agreed within the errors.

\begin{figure*}[htbp]
\centering
  \includegraphics*[width=120mm]{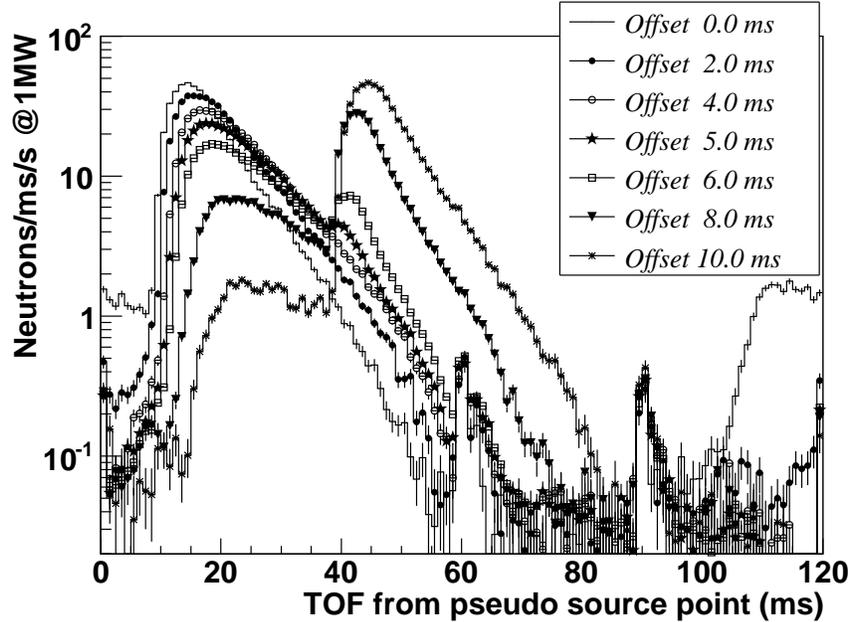}
  \caption{TOF spectra of Doppler-shifted neutrons with different phase offsets.
  Each time origin was set to the time when the Doppler mirror holder passed the pseudo source point. \label{fig010}}
\end{figure*}

The TOF spectra of the Doppler-shifted neutrons from the pseudo source point with phase offsets of 0.0, 2.0, 4.0, 5.0, 6.0, 8.0, and $10.0~\mathrm{ms}$ are shown in FIG. \ref{fig010}.
The timing origins of the horizontal axis for the TOF spectra were adjusted to each phase offset.
The origin of the offset was determined at the timing when the photo-sensor pulse coincided with the $25/3$-Hz timing signal.

The TOF spectra with offsets of 0.0--$6.0~\mathrm{ms}$ have peaks at $15~\mathrm{ms}$.
The peak height begins to decrease with increasing phase offset, and a peak at $45~\mathrm{ms}$ arises.
This is because the four dips in FIG. \ref{fig009}, which imply neutron reflection by the Doppler mirror, move to the right in response to the increased phase offset, as shown in the lower part of the figure.
Therefore, the neutron reflection timing gradually becomes mismatched with any given VCN pulse, but then begins to match with the next pulse.
The timing relation between the dips and VCN pulses in FIG. \ref{fig009} shows that the spectra with 0.0- and 10.0-ms offsets differ in this way only, but are otherwise intrinsically identical.
Note that, in these measurements, the length of the UCN extraction guide was not $143~\mathrm{mm}$, but $121~\mathrm{mm}$.

The neutron backgrounds were evaluated through measurement with the off-position of the Doppler mirror from the beamline and were then subtracted from each spectrum.
The background count rate was $2.8~\mathrm{cps}$ with a beampower equivalent to $1~\mathrm{MW}$, while that of the Doppler-shifted neutrons for the 5.0-ms offset was $4.1 \times 10^{2}~\mathrm{cps}$.
Hence the background was negligibly small.
The small peaks that appear at 60, 90, and $120~\mathrm{ms}$ in the figure represent the background neutrons scattered by the mirror holder; this can be concluded based on the fact that the small peaks appear immediately after the mirror crosses the VCN pulse.
It is very difficult to block these background neutrons in the detector arrangement, because they enter the UCN extraction guide with the UCN.
Although the background neutrons can be evaluated precisely by utilizing plain silicon wafers instead of the Doppler mirror,
the Doppler mirror can not be dismounted from the holder easily.

The neutron yield between 31 and $35~\mathrm{ms}$, for which $v$ was sufficiently less than that of the UCN and the spectrum did not overlap the peak from next pulse,
was at a maximum for a phase offset of 5.0 ms.
Therefore, the long-term experiments discussed below were performed with a $5.0$-$\mathrm{ms}$ phase offset.

\begin{figure*}[htbp]
  \centering
  \includegraphics*[width=120mm]{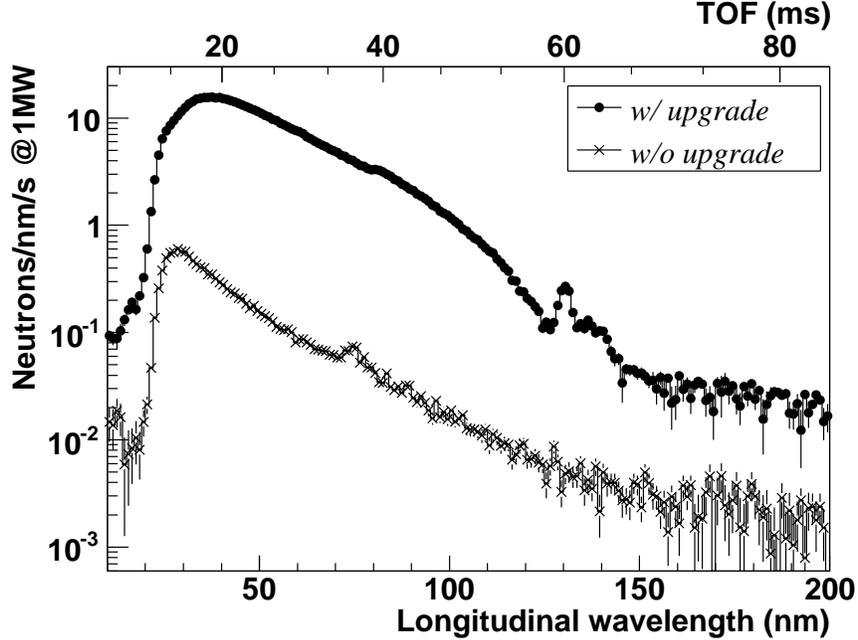}
  \caption{Spectra of Doppler shifted neutrons.
  The closed circles show the result with the upgrades, while the crossed symbols show a demonstration experiment result (before the upgrade).\label{fig011}}
\end{figure*}

The longitudinal wavelength spectrum of the Doppler-shifted neutrons with 5.0-ms phase offset is shown in FIG. \ref{fig011},
along with that of the demonstration experiment.
The wavelength was calculated based on the TOF distance and the duration between the offset and detection times, while
considering the deceleration due to gravity.
The flight path in the detector chamber was not considered in this figure.
Each data set was compensated against the UCN detector efficiency given in equation (\ref{equ08}), while the attenuation in the Al window was ignored.
The background values were subtracted from each plot.
The small peak at $80~\mathrm{nm}$ represents the abovementioned second peak arising from the reflection of the next VCN pulse.
The peak that appears at $130~\mathrm{nm}$ is due to the background neutrons scattered by the Doppler mirror holder.
The $1~\mathrm{MW}$-equivalent total output was $6.0 \times 10^{2}~\mathrm{cps}$.
Note that the phase offset of the demonstration experiment was estimated as $4.5~\mathrm{ms}$, from the VCN dips,
and the $1~\mathrm{MW}$-equivalent total count rate was $13~\mathrm{cps}$.

The part of the spectrum corresponding to wavelengths longer than $123~\mathrm{nm}$, which is the cutoff wavelength of the Al window~\cite{lec33}, was regarded as a constant background.
This background was most likely due to the neutrons remaining in the vacuum chamber, which would have been scattered by the rotating blade or the mirror holder and would have then entered the detector through the UCN extraction guide.
These suspect neutrons surely mixed in the UCN region; thus, we assumed that a constant background with the same height as the $124~\mathrm{nm}$ count increased the overall UCN spectrum.
The constant background count between 58 and $123~\mathrm{nm}$ was estimated as $7.8~\mathrm{cps}$.
The total count of the second peak was estimated to be $2.1~\mathrm{cps}$; this estimation was conducted by approximating the peak as triangular with linear slope.
Accordingly, the $1~\mathrm{MW}$-equivalent count rate for neutrons with longitudinal wavelengths between 58 and $123~\mathrm{nm}$ was estimated to be $1.6\times 10^{2}~\mathrm{cps}$.
In the demonstration experiment~\cite{lec19}, the neutron count calculated in the same manner was $1.8~\mathrm{cps}$.
Thus, the count rate increased 91-fold from demonstration experiment.

\subsection{\label{sec42}Comparison with simulation}
\begin{figure*}[htb!p]
  \centering
  \includegraphics*[width=80mm]{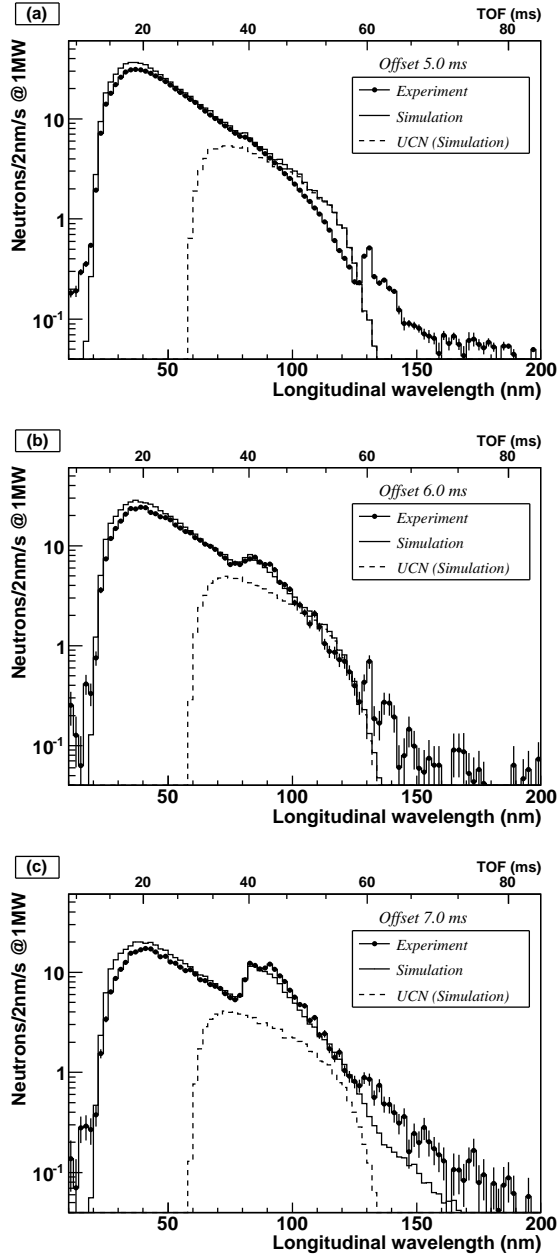}
    \caption{Comparison of simulated and experimental results of longitudinal UCN wavelength spectra
    at phase offsets of (a--c) 5.0, 6.0, and $7.0~\mathrm{ms}$, respectively.
    The dotted, solid, and dashed lines indicate measured data, simulated spectra, and the simulated true UCN spectra, respectively.
    The UCN simulation results are normalized by the measured VCN amount between 2.5 and $6.0~\mathrm{nm}$.
  \label{fig012}}
\end{figure*}

The measured and simulated spectra of the Doppler-shifted neutrons at the UCN detector position with 5.0-, 6.0-, and 7.0-ms phase offset are shown in parts (a--c) of FIG. \ref{fig012}, respectively.
The dashed lines indicate neutrons with wavelength longer than $58~\mathrm{nm}$ (true UCN) calculated via the UCN simulation.
The measured spectra were compensated based on the UCN detector efficiency given in equation (\ref{equ08}) only.

In the UCN simulation, the VCNs were seeded at the unpolarized branch window in the same manner as described in section \ref{sec32}.
Only the VCNs with 120--155-$\mathrm{m/s}$ velocity were calculated, in order to reduce the computation time, under the condition that the first and second peaks were fully produced.
The reflectivity of the Doppler mirror was taken from that of the stacked mirrors in FIG. \ref{fig008}, and the mirror size was set to $30~\mathrm{mm} \times 30~\mathrm{mm}$.
Scattering of the VCNs by the mirror holder was ignored.
The neutron path length in the detector gas chamber was calculated using equation (\ref{equ08}), up to the detector length of $50~\mathrm{mm}$.
The longitudinal wavelength was calculated by assuming the TOF distance was $199~\mathrm{mm}$, while the TOF was the sum of the flight times inside and outside the detector, as in FIG. \ref{fig011}.
The time origin of the TOF was set to the moment when the Doppler mirror passed through the pseudo source point.
The detection efficiency was calculated using the true neutron wavelength, which was computed from the magnitude of the velocity vector in the UCN simulation.
One detection event was weighted by this efficiency and then corrected in the same way as in the experimental data analysis.

As can be seen in FIG. \ref{fig012} (a), for the 5.0-$\mathrm{ms}$ offset, the simulated (solid line) and measured (dotted line) spectrum shapes are in good agreement for wavelengths less than $94~\mathrm{nm}$.
Disagreement occurs for wavelengths longer than this value, which is assumed to be caused by the metal beams supporting the foil.
The measured count in the UCN region without the second peak, between 58 and $78~\mathrm{nm}$, was $6.5\%$ less than the simulated one.
By considering the velocity divergence of the UCN, the simulated true UCN count (dashed line) around 60 to $70~\mathrm{nm}$ was significantly smaller than the measured longitudinal UCN count. This was
because Doppler-shifted neutrons with wide divergence were collected by the UCN extraction guide and, hence,
many neutrons around the longitudinal UCN limit were faster than $6.8~\mathrm{m/s}$.

By normalizing the total count of the simulated spectrum by the measured result, the dashed line in FIG. \ref{fig012} (a) is scaled to the true UCN spectrum in the experiment.
In this case, the UCN count rate was estimated to be $92~\mathrm{cps}$.
However, this value is slightly larger than the true UCN count in the experiment, because the measured spectrum had the constant background mentioned in section \ref{sec41}
and the spectrum around the cutoff wavelength of Al was smaller than the simulated one.
The constant background in the true UCN region and the disagreement around the cutoff wavelength were estimated to be $7.8$ and $3.6~\mathrm{cps}$, respectively.
After the removal of the counts from $92~\mathrm{cps}$, the true UCN count rate in the experiment was estimated to be $80~\mathrm{cps}$.
This is half of the measured longitudinal UCN count.
As a result, the UCN count rate increased 45-fold against the demonstration experiment.
This is approximately the same as the VCN increase at the pseudo source point.

The shape of the simulated spectra for the 6.0- and 7.0-ms offsets are also well matched to the experimental results, including the occurrence of the second peak, as shown in FIG. \ref{fig012} (b) and (c).
In the same manner as the 5.0-ms offset, the true UCN count rates for the 6.0- and 7.0-ms offsets were estimated to be 72 and $60~\mathrm{cps}$.
The simulated first peak for the 7.0-ms offset case was $7.2\%$ larger and the second peak was $24\%$ smaller than the experiment.
As the discrepancy was minimized for the 7.3-$\mathrm{ms}$ offset time setting in the UCN simulation, the TOF distance of the incident beam was $4~\mathrm{cm}$ shorter or the reflectivity distribution of the Doppler mirror was $1\%$ higher than our estimation.
The systematic error of the phase offset measured by the photo-sensor could be also a source of the discrepancy.

\section{\label{sec5}UCN density estimation}
The maximum phase space density of the UCNs is limited to that of the incident beam,
because a conservative force such as those acting in the Doppler shifter or the gravity does not cause a change in phase space density.
Therefore, in this section, we estimate the phase space densities of the VCN beam and the produced UCNs,
along with the number density of the UCNs at production in the same manner as Refs.~\cite{lec10,lec19}.

\begin{figure*}[htbp]
  \centering
  \includegraphics*[width=90mm]{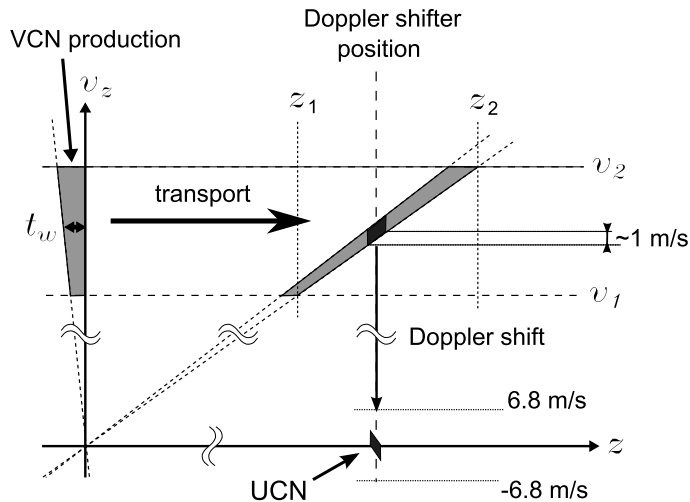}
  \caption{Schematic diagram of phase space distribution of pulsed beam in $z$-$v_{z}$ plane.\label{fig013}}
\end{figure*}

The phase space density of the pulsed neutrons is calculated as follows:
if a group of neutrons with various velocities is produced in a neutron source, the phase space distribution within a certain velocity range in the $z$-$v_{z}$ plane is expressed simply as a trapezoid-shaped distribution (see FIG. \ref{fig013}). 
Here, $v_{2}$ and $v_{1}$ are upper and lower velocity limits, respectively, $z_{2}$ and $z_{1}$ are the head and tail of the neutron pulse position, respectively, and $t_{w}$ is the time width of the neutron pulse.
In the TOF measurement, the number of neutrons ($N$) between $z_{1}$ and $z_{2}$ is ordinarily regarded as being those between $v_{1}$ and $v_{2}$.

If the repetition frequency of the pulse is $f$, the duration of the beam per $1~\mathrm{s}$ is $ft_{w}$.
Therefore, the peak phase space density inside the trapezoid of the pulsed beam is $1/ft_{w}$ times greater than that of a continuous beam with the same average flux.
The spatial volume of the beam passing through a unit area of the neutron detector per unit time is approximately described as the mean velocity $v=(v_{1} + v_{2})/2$,
if $(v_{2} - v_{1})$ is sufficiently small.
In this case, $N$ represents the average flux of the pulsed beam.
In addition, the divergences $\Delta v_{x}$ of $v_{x}$ and $\Delta v_{y}$ of $v_{y}$ 
are used to calculate the velocity phase space volume.
Accordingly, the peak phase space density, $\rho(v)$, is expressed as 
\begin{equation}
       \rho(v) = N / (ft_{w} \cdot v \cdot \Delta v_{x} \cdot \Delta v_{y} \cdot (v_{2}-v_{1})).
       \label{equ09}
\end{equation}

\subsection{\label{sec51} Phase space density of VCN beam}
The parameters used in the equation (\ref{equ09}) and calculated peak phase space densities at the unpolarized branch window
and the pseudo source point are summarized in TABLE. \ref{tab02}.
The common parameters were estimated as follows.
The pulse width (FWHM) was estimated for 136-$\mathrm{m/s}$ VCN by fitting the time spectra of the official moderator simulation for the BL05 port~\cite{lec28}.
The repetition frequency of the pulsed beam was as mentioned in section \ref{sec3}.
The parameters at the unpolarized branch window were estimated as follows.
The $1~\mathrm{MW}$-equivalent flux between $134~\mathrm{m/s}$ ($2.95~\mathrm{nm}$) and $138~\mathrm{m/s}$ ($2.87~\mathrm{nm}$), was calculated from equation (\ref{equ06}) without the constant $C_{ b}$.
The FWHMs of VCN divergences mentioned in section \ref{sec31} were selected as the $x'$- and $y'$-divergence.
The divergences were converted into the velocity widths of $136~\mathrm{m/s}$-VCN, namely $\Delta v_{x'}$ and $\Delta v_{y'}$.
The ratio of VCNs in the $\Delta v_{x'}$- and $\Delta v_{y'}$-region to total VCNs in the velocity phase space was named ``selected neutron yield".
The flux multiplied by the selected neutron yield was substituted for $N$ in equation (\ref{equ09}).
The parameters at the pseudo source point were estimated as follows.
The $1~\mathrm{MW}$-equivalent flux that passes the cross section of $30~\mathrm{mm} \times 30~\mathrm{mm}$ was estimated from the UCN simulation and the data set at the square collimator on the VCN exit window as mentioned in section \ref{sec32}.
The $x$- and $y$-divergence of VCN were determined by the boundaries of dense region in histograms of calculated beam divergence.
The divergences were converted into $\Delta v_{x}$ and $\Delta v_{y}$.
The meaning of ``selected neutron yield" and its usage are the same as above mentioned.

\begin{table}[htbp]
  \caption{
  Parameters used in the calculation of peak phase space density and the result of the calculation at the unpolarized branch window and the pseudo source point.
  $x'$-, $y'$-, $x$- and $y$-divergence were converted into the velocity widths of $\Delta v_{x'}$, $\Delta v_{y'}$, $\Delta v_{x}$ and $\Delta v_{y}$, respectively.
  ``Selected neutron yield" means the ratio of VCNs within the $\Delta v_{x'}$- (or $\Delta v_{x}$-) and $\Delta v_{y'}$- (or $\Delta v_{y}$-) region to total VCNs in the velocity phase space.
  }
  \label{tab02}
  \begin{center}
    \begin{tabular}{p{7.5cm} p{7.5cm}}
      \hline\hline
      \multicolumn{2}{l}{Common parameters}~~\\
      \hline
      Pulse width (FWHM)~~& $4.9 \times 10^{2}~\mathrm{\mu s}$~~\\
      Repetition frequency~~& $25~\mathrm{Hz}$~~\\
      \hline\hline
      \multicolumn{2}{l}{Parameters at the unpolarized branch window}~~\\
      \hline
      Neutron flux in 134--$138~\mathrm{m/s}$~~& $9.6 \times 10^{4}~\mathrm{cm^{-2}s^{-1}}$~~\\
      $x'$-divergence (FWHM)~~& $50~\mathrm{mrad}$~~\\
      $y'$-divergence (FWHM)~~& $54~\mathrm{mrad}$~~\\
      $\Delta v_{x'}$~~& $6.8~\mathrm{m/s}$~~\\
      $\Delta v_{y'}$~~& $7.3~\mathrm{m/s}$~~\\
      Selected neutron yield~~& $61\%$~~\\
      \hline
      Phase space density~~& $1.8~\mathrm{cm^{-3}(m/s)^{-3}}$~~\\
      \hline\hline
      \multicolumn{2}{l}{Parameters at the pseudo source point}~~\\
      \hline
      Neutron flux in 134--$138~\mathrm{m/s}$~~& $3.5 \times 10^{4}~\mathrm{cm^{-2}s^{-1}}$~~\\
      $x$-divergence~~& $0 \pm 24~\mathrm{mrad}$~~\\
      $y$-divergence~~& $0 \pm 70~\mathrm{mrad}$~~\\
      $\Delta v_{x}$~~& $6.5~\mathrm{m/s}$~~\\
      $\Delta v_{y}$~~& $19~\mathrm{m/s}$~~\\
      Selected neutron yield~~& $63\%$~~\\
      \hline
      Phase space density~~& $0.26~\mathrm{cm^{-3}(m/s)^{-3}}$~~\\
      \hline
    \end{tabular} 
  \end{center}
\end{table}

As shown in TABLE. \ref{tab02}, the phase space density at the pseudo source point is $15\%$ of that at the unpolarized branch.
However, the value is more than 3 times larger than the UCN phase space density of $8.4 \times 10^{-2}~\mathrm{neutrons / cm^3 / (m/s)^3}$ at ILL/PF2~\cite{lec09}.
If the flux at the unpolarized branch window is ideally transported to the pseudo source point, the efficiency is calculated to be $76\%$ from TABLE. \ref{tab01}.
Hence, the flux at the pseudo source point was expected to be $7.4 \times 10^{4}~\mathrm{neutrons / cm^2 / s}$.
However, the measured flux, and accordingly measured phase space density, was half the expected value.

Taking the reflectivity of the Doppler mirror as being 0.229, the phase space density of the reflected VCN is expected to be $6.0 \times 10^{-2}~\mathrm{neutrons / cm^3 / (m/s)^3}$.
The value is nearly the same as the phase space density of UCN at ILL/PF2.
If the UCNs fill the entire spherical velocity volume with $6.8$-$\mathrm{m/s}$ radius,
the UCN number density is calculated as being $80~\mathrm{UCN/cm^{3}}$.

\subsection{\label{sec52} UCN phase space and number densities}
The phase space density of the detected UCN was estimated as follows.
The $1~\mathrm{MW}$-equivalent measured UCN count was estimated to be $9.6~\mathrm{UCN/pulse}$ from the conclusion of section \ref{sec42}.
The spatial volume where the UCNs were produced was $3.0~\mathrm{cm} \times 3.0~\mathrm{cm} \times 6.0~\mathrm{cm}$ according to the UCN simulation.
The velocity phase space in which the UCNs were produced has a partial spherical shell shape with radius from 3.7--$7.1~\mathrm{m/s}$, determined by considering the Al cutoff and gravity.
The volume of the phase space was numerically calculated to be $163~\mathrm{(m/s)^3}$.
Therefore, the phase space density of the UCN at production is $1.1 \times 10^{-3}~\mathrm{neutrons / cm^3 / (m/s)^3}$, which is $1.8\%$ of the phase space density of the VCN reflected by the Doppler mirror.
This corresponds to $1.4~\mathrm{UCN / cm^3}$, assuming a spherical velocity phase space volume of 6.8-$\mathrm{m/s}$ radius.
The density is 12 times larger than that produced at the Argonne National Laboratory~\cite{lec11} and is as high as that in the nEDM experiment at ILL/PF2
which set the current upper limit of $2.9 \times 10^{-26}~\mathrm{e \cdot cm}$ (90\% C.L.)~\cite{lec02,lec03}.

The source of the reduction in the phase space density can be explained by an overestimation of the phase space volume because of empty components contained within it.
During the UCN production, the Doppler mirror overlaps the injected VCN beam for only $0.9~\mathrm{ms}$, as mentioned in section \ref{sec2}, so the range of the reflected VCN velocity in the z-direction is $1~\mathrm{m/s}$.
The UCN $v_{z}$ range is identical to that of the VCN expressed as a black parallelogram in FIG. \ref{fig013},
if we assume a Doppler mirror moving parallel to the VCN for simplicity.

In the UCN simulation, the FWHM of the $v_{z}$ distribution of the reflected VCNs was $0.58~\mathrm{m/s}$.
Therefore, the velocity phase space occupied by the UCN has a disk shape with 0.58-$\mathrm{m/s}$ thickness and 6.8-$\mathrm{m/s}$ radius.
The volume of the disk is $84.9~\mathrm{(m/s)^{3}}$, which is $6.4\%$ that of a sphere with radius 6.8 $\mathrm{m/s}$.
By multiplying the VCN phase space density obtained in section \ref{sec51} and the velocity volume of the disk together, the UCN density is calculated as $5.1~\mathrm{UCN/cm^{3}}$.
The spatial volume was also overestimated because the UCN density in the volume was not uniform.
In the UCN simulation, the number density of UCNs in the 1-$\mathrm{cm^{3}}$ cube at the psuedo source point just after production was $3.9~\mathrm{UCN / cm^3}$,
while the average density in $3.0~\mathrm{cm} \times 3.0~\mathrm{cm} \times 6.0~\mathrm{cm}$-volume was $1.2~\mathrm{UCN / cm^3}$.
Thus, the overestimations in the spatial and velocity volumes roughly explain the reduction of the phase space density of the UCNs.

\begin{figure*}[htbp]
  \centering
  \includegraphics*[width=100mm]{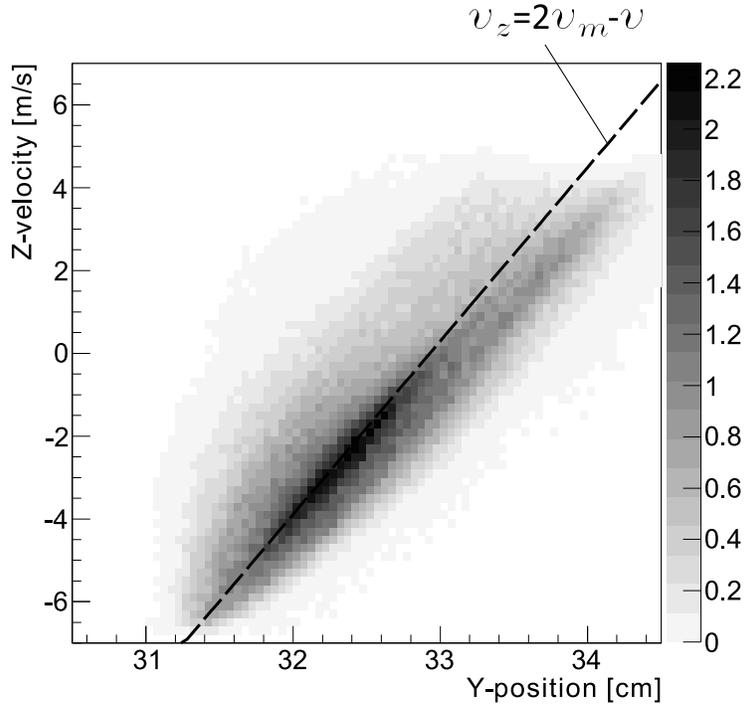}
  \caption{UCN distribution in $y$-$v_{z}$ plane as calculated by the UCN simulation.
  The dashed line shows the dominant term of equation (\ref{equ05}) in the region, that is $(2 v_{m} - v)$ where $v = 17.2/0.125 \simeq 138~\mathrm{m/s}$ and $v_{m} = (200 \pi / 3) y$.\label{fig014}}
\end{figure*}

As the UCNs produced by the Doppler shifter have a position and velocity correlation, they are concentrated in a part of the phase space volume.
As described in equation (\ref{equ05}), the Doppler-shifted neutrons are distributed on a line of $(2v_{m} - v)$ in the $y$-$v_{z}$ plane, as shown in FIG. \ref{fig014}, because of the difference of the rotating radius.
Therefore, if we take the phase space volume as a product of the spatial and velocity volume with which the UCNs are produced, the effective phase space density becomes small.
If we use a parallel moving Doppler mirror, such as an excentric double motor~\cite{lec34},
or modify the UCN velocity with a magnetic accelerator system like the UCN rebuncher, we can obtain a UCN beam with smaller $v_{z}$ divergence.

\section{\label{sec6}Conclusion and future outlook}
We have developed the proposed Doppler shifter for the first UCN source at J-PARC and significantly improved the UCN output by replacing or adding neutron optical components for practical use.
The Doppler-shifted neutrons have an 8.33-Hz sharply pulsed time structure of 4.4-ms width.
The $1~\mathrm{MW}$-equivalent neutron count rate for neutrons of longitudinal wavelength between 58 and $123~\mathrm{nm}$ is $1.6 \times 10^{2}~\mathrm{cps}$.
The true UCN count rate is estimated to be $80~\mathrm{cps}$, according to a Monte Carlo simulation.
The UCN number density is estimated to be at least $1.4~\mathrm{UCN/cm^{3}}$ at production.
The Doppler shifter works almost as designed and produces sufficient UCN output for practical use. 

\begin{figure*}[htbp]
  \centering
  \includegraphics*[width=120mm]{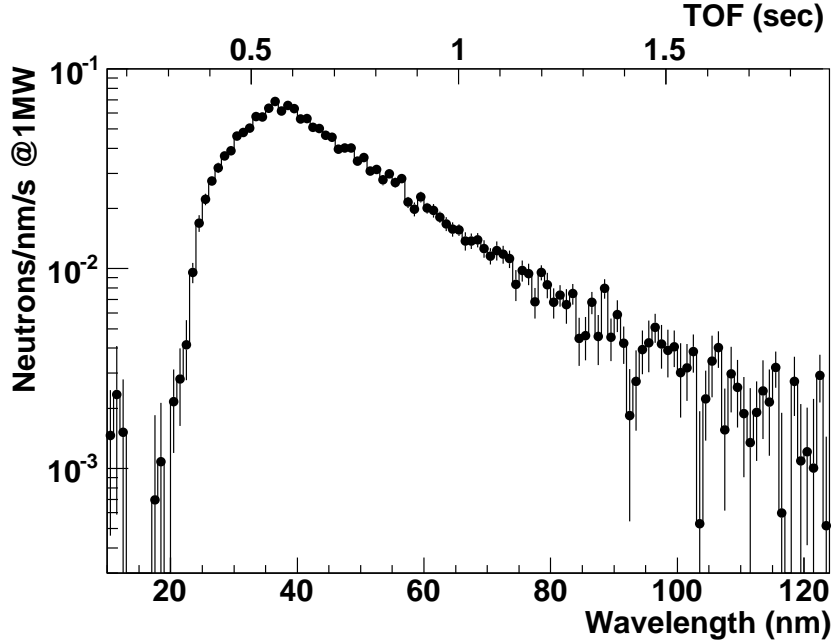}
  \caption{TOF spectrum of UCN transport experiment with Ni-coated guides of $5.6~\mathrm{m}$ in length.
  The UCN pulses were thinned to $0.52~\mathrm{Hz}$.
  \label{fig015}}
\end{figure*}

Through application of the Doppler shifter, we have already begun R \& D of apparatus for use in our search for the nEDM~\cite{lec20}.
The result of the UCN transport experiment with the 5.6 m-long guide is shown in FIG. \ref{fig015}, as an example.
UCNs were guided to the horizontal direction with an L-shape mirror guide connected to the 5.6 m-long guide.
UCN count was reduced to $1/3$ in this section.
The mean reflectivity of the 5.6 m-long guide was estimated to be $97\%$.
The 1 MW-equivalent total count rate increased to $1.9~\mathrm{cps}$ by the improvement of the Doppler shifter and the UCN region was $0.39~\mathrm{cps}$,
although the UCN pulses were thinned to $1 / 16$, $0.52~\mathrm{Hz}$, with the shutter.
As the result of the improvement, the Doppler shifter allows us to perform the UCN rebuncher experiment at BL05 in a practical time span.
At present, we are planning the performance test of the improved UCN rebuncher with an experimental setup incorporating the UCN guides, along with
the development of a UCN reflectometer and storage experiments.

\section{\label{sec7}Acknowledgements}
This research was partially supported by Ministry of Education, Science, Sports, and Culture Grants-in-Aid for Scientific Research (A) 23244047 and
(A) 26247035, and by the JSPS, Grant No. 23360428.
The Doppler shifter experiment was approved by the Neutron Scattering Program Advisory Committee of IMSS, KEK (Proposal No. 2009S03 and 2014S03).
The neutron reflectivity experiment was approved by the Neutron Science Proposal
Review Committee of J-PARC/MLF (Proposal No. 2009S08 and 2014B0263) and supported by the Inter-University Research Program on Neutron Scattering of IMSS, KEK.

\section{\label{sec8}References}


\begin{thebibliography}{34}

\bibitem{lec01} N. F. Ramsey, Rep. Prog. Phys. {\bf 45}, 95 (1982).%1
\bibitem{lec02} P. G. Harris et al., Phys. Rev. Lett. {\bf 82}, 904 (1999).%2
\bibitem{lec03} C. A. Baker et al., Phys. Rev. Lett. {\bf 97}, 131801 (2006).%3
\bibitem{lec04} K.-J. K\"{u}gler, K. Moritz, W. Paul, and U. Trinks, Nucl. Instr. and Meth. Phys. Res. Sect. A {\bf 228}, 240 (1985).%4
\bibitem{lec05} A. Serebrov et al., Phys. Lett. B {\bf 605}, 72 (2005).%5
\bibitem{lec06} G. J. Mathews, T. Kajino, and T. Shima, Phys. Rev. D {\bf 71}, 021302(R) (2005).%6
\bibitem{lec07} B. D. Fields and S. Sarkar, Phys. Lett. B {\bf 667}, 228 (2008).%7
\bibitem{lec08} R. Golub, Rev. Mod. Phys. {\bf 68}, 329 (1996). %8
\bibitem{lec09} A. Steyerl et al., Phys. Lett. A {\bf 116}, 347 (1986).%9
\bibitem{lec10} T. W. Dombeck et al., Nucl. Instr. and Meth. {\bf 165}, 139 (1979).%10
\bibitem{lec11} T. O. Brun et al., Phys. Lett. A {\bf 75}, 223 (1980).%11
\bibitem{lec12} S. Mayer et al., Nucl. Instr. and Meth. Phys. Res. Sect. A {\bf 608}, 434 (2009).%12
\bibitem{lec13} J. W. Lynn et al., Physica B {\bf 120}, 114 (1983).%13
\bibitem{lec14} K. Mishima et al., Nucl. Instr. and Meth. Phys. Res. Sect. A {\bf 600}, 342 (2009).%14
\bibitem{lec15} Y. Arimoto et al., Prog. Theor. Exp. Phys. {\bf 2012}, 02B007 (2012).%15
\bibitem{lec16} M. Hino, M. Kitaguchi, and Y. Kawabata, Kurri Prog. Rep. 2009, Kyoto Univ., 146 (2010).%16
\bibitem{lec17} M. Hino et al., Nucl. Instr. and Meth. Phys. Res. Sect. A {\bf 529}, 54 (2004).%17
\bibitem{lec18} M. Hino et al., Nucl. Instr. and Meth. Phys. Res. Sect. A {\bf 600}, 207 (2009).%18
\bibitem{lec19} K. Mishima et al., J. Phys.: Conf. Ser. {\bf 528}, 012030 (2014).%19
\bibitem{lec20} Proposal to J-PARC, ~{}http:{}\slash{}\slash{}j-parc.jp{}\slash{}researcher{}\slash{}Hadron{}\slash{}en{}\slash{}pac{}\_{}1001\slash{}pdf\slash{}KEK{}\_{}J-PARC-PAC2009-11.pdf.%20
\bibitem{lec21} Y. Arimoto et al., Phys. Rev. A {\bf 86}, 023843 (2012).%21
\bibitem{lec22} B. Alefeld, G. Badurek, and H. Rauch, Z. Phys. B {\bf 41}, 231 (1981).%22
\bibitem{lec23} S. V. Grigoriev, A. I. Okorokov, and V. V. Runov, Nucl. Instr. and Meth. Phys. Res. Sect. A {\bf 384}, 451 (1997).%23
\bibitem{lec24} T. Ino, H. Otono, K. Mishima and T. Yamada., J. Phys.: Conf. Ser. {\bf 528}, 012039 (2014).%24
\bibitem{lec25} K. Hirota et al., Phys. Chem. Chem. Phys. {\bf 7}, 1836 (2005).%25
\bibitem{lec26} M. Katagiri et al., Nucl. Instr. and Meth. Phys. Res. Sect. A {\bf 529}, 274 (2004).%26
\bibitem{lec27} Y. Ikeda, J. Nucl. Mater. {\bf 343}, 7 (2005).%27
\bibitem{lec28} Technical Details, Materials and Life Science Experimental Facility, ~{}http{}\slash{}\slash{}j-parc.jp{}\slash{}researcher{}\slash{}MatLife{}\slash{}en{}\slash{}Instrumentation{}\slash{}ns3.html.%28
\bibitem{lec29} N. L. Yamada et al., Eur. Phys. J. Plus {\bf 126}, 108 (2011).%29
\bibitem{lec30} K. Mitamura et al., Polymer J. {\bf 45}, 100 (2013).%30
\bibitem{lec31} A. Steyerl and H. Vonach, Z. Phys. A {\bf 250}, 166 (1972).%31
\bibitem{lec32} A.-J. Dianoux, G. Lander (Eds.), ILL Neutron Data Booklet, Institut Laue-Langevin, pp. 1.1--10 (2003).%32
\bibitem{lec33} V. F. Sears, Neutron News {\bf 3}, 26 (1992).%33
\bibitem{lec34} P. Fierlinger, Physics Procedia {\bf 51}, 102 (2014).%34

\end{thebibliography}
\end{document}